\documentclass[twocolumn,eqsecnum,showpacs,showkeys,amsmath,amssymb]{revtex4}
\usepackage{graphicx}
\usepackage{bm}

\def\vec#1{\mathchoice{\mbox{\boldmath$\displaystyle#1$}}
{\mbox{\boldmath$\textstyle#1$}}
{\mbox{\boldmath$\scriptstyle#1$}}
{\mbox{\boldmath$\scriptscriptstyle#1$}}}
\makeatletter
\newcommand\erfc{\mathop{\operator@font erfc}\nolimits}
\def\slashchar#1{\setbox0=\hbox{$#1$}
   \dimen0=\wd0 \setbox1=\hbox{/} \dimen1=\wd1
   \ifdim\dimen0>\dimen1 \rlap{\hbox to \dimen0{\hfil/\hfil}} #1
   \else  \rlap{\hbox to \dimen1{\hfil$#1$\hfil}} / \fi}
\def\p{\slashchar{p}}
\def\w{\omega} 

\makeatother

\begin{document}
 
\title{Spectral quark model and low-energy hadron phenomenology}

\author{Enrique Ruiz Arriola}
\email{earriola@ugr.es}
\affiliation{Departamento de F\'{\i}sica Moderna, Universidad de
Granada, E-18071 Granada, Spain}
\author{Wojciech Broniowski} 
\email{Wojciech.Broniowski@ifj.edu.pl} 
\affiliation{The H. Niewodnicza\'nski Institute of Nuclear Physics,
PL-31342 Krak\'ow, Poland}
\date{21 January 2003}

\begin{abstract}
We propose a spectral quark model which can be applied to low
energy hadronic physics. The approach is based on a generalization 
of the Lehmann representation of the quark propagator. We work at the
one-quark-loop level. Electromagnetic and chiral invariance are
ensured with help of the gauge technique which provides particular
solutions to the Ward-Takahashi identities. General
conditions on the quark spectral function follow from natural
physical requirements. In particular, the function is normalized, its all positive 
moments must vanish, while the physical observables depend on
negative moments and the so-called log-moments. As a consequence, 
the model is made finite, dispersion relations hold, chiral anomalies
are preserved,  and the twist expansion is free from logarithmic
scaling violations, as requested of a low-energy model. We study a
variety of processes and show that the framework is very simple and
practical. Finally, incorporating the idea of vector-meson
dominance, we present an explicit construction of the quark
spectral function which satisfies all the requirements. The
corresponding momentum representation of the resulting quark propagator
exhibits only cuts on the physical axis, with no poles present
anywhere in the complex momentum space. The
momentum-dependent quark mass compares very well to recent lattice
calculations. A large number of predictions and relations, valid at
the low-energy scale of the model, can be deduced from our
approach for such quantities as the pion light-cone wave function, non-local quark
condensate, pion transition form factor, pion valence parton distribution
function, {\em etc.} These quantities, obtained at a low-energy scale of the model,  
have correct properties, as requested by symmetries and anomalies. They also have 
pure twist expansion, free of logarithmic corrections, as requested by the QCD 
factorization property.
\end{abstract}



\maketitle 

\section{Introduction \label{sec:intro}} 
The term {\em chiral quark model} has become a generic name for any
relativistic field
 theory aiming at the description of the
non-perturbative features of QCD.  Numerous approaches (for reviews
see, {\em e.g.},
\cite{NJLrev:reg,NJLrev:klev,NJLrev:volkov,NJLrev:jap,NJLrev:Bo,NJLrev:Tue,NJLrev:ripka,Ru02},
and references therein) share a number of common features. Firstly,
they incorporate dynamical quarks as the only explicit degrees of
freedom. Secondly, they provide particular solutions to the chiral and
electromagnetic Ward-Takahashi identities. Although there is no doubt
that chiral quark models provide a reasonably accurate quantitative
description of hadronic properties, there is a lack of systematics in
the construction of any particular dynamical model. A source of
ambiguity is the fact that chiral quark models are, supposedly, an
approximation to the low-energy non-perturbative QCD dynamics. In
this regard, an essential ingredient is the introduction of  a
practical suppression of high-energy degrees of freedom,
necessary in order to separate the low-energy regime, where the
model is supposed to work, and the high-energy regime, where the
genuine QCD dynamics, in terms of explicit quarks and gluons, should
set in. This defines a certain scale, or cut-off, which acquires a
physical meaning and which should be kept throughout the
calculation. The precise way how this high-energy cut-off should be
introduced is not at all clear, although some constraints can be
imposed on the basis of the relativistic and gauge invariance. 
For a discussions of various popular approaches and associated 
problems see, {\em e.g.}, \cite{Ru02}.

In this work we introduce a novel approach, the {\em spectral regularization} of the 
chiral quark model, 
 based on the formal introduction of the Lehmann
representation \cite{IZ80} for
 the quark propagator,
\begin{eqnarray}
S({p}) = \int_C d \omega { \rho( \omega ) \over
\slashchar{p} - \omega  } ,
\label{eq:spec0} 
\end{eqnarray}
where $\rho(\omega)$ is the spectral function and $C$ denotes a contour in the
complex $\omega$ plane chosen in a suitable way.  As will become clear
in Sect. \ref{sec:gauge}, the spectral regularization allows to
explicitly solve the chiral and electromagnetic Ward-Takahashi identities in a
rather simple manner, through the use of the so-called {\em gauge
technique}~\cite{DW77,De79}. The method has already been sketched in a
previous work by one of us~\cite{Ru01}. Of course, any solution of the
Ward-Takahashi identities cannot be complete due to the existence of
transverse terms, which necessarily appear in the underlying theory,
and which can only be uniquely determined in QCD. In fact, any
possible realization of a chiral quark model, if properly regularized,
represents in a sense a particular solution to the Ward-Takahashi
identities. This point is not fully appreciated. In
Sect.~\ref{sec:gauge} we will provide a minimal set of solutions to the
Ward-Takahashi identities and study their consequences.

Throughout the paper the model is considered at the one-quark-loop level
and in the chiral limit of the vanishing current quark masses. Therefore
our present predictions are made at the leading-$N_c$ level and in the chiral limit.
 
 The proper normalization and the conditions of
finiteness of hadronic
 observables are achieved by requesting what
we call the {\em spectral conditions} 
 for the moments of the quark
spectral function, $\rho(\omega)$, namely
\begin{eqnarray}
\rho_0 && \equiv \int d\omega \rho(w) = 1, \label{rho0} \\
\rho_n && \equiv \int d\omega \omega^n \rho(\omega) = 0, 
 \;\;\; {\rm for} \; n=1,2,3,... \label{rhon}
\end{eqnarray} 
It will be shown that the physical observables are proportional to the inverse moments, 
\begin{eqnarray}
\rho_{-k} && \equiv \int d\omega \omega^{-k} \rho(\omega), 
 \;\;\; {\rm for} \; k=1, 2, 3,... \label{rhoinv}
\end{eqnarray} 
as well as to the {\em ``log moments''},
\begin{eqnarray}
&& \rho'_n\equiv \int d\omega \log(\omega^2/\mu^2) \omega^n \rho(\omega) \nonumber \\
&& =\int d\omega \log(\omega^2) \omega^n \rho(\omega), \;\;\; {\rm for} \; n=2,3,4,...  
\label{rholog}
\end{eqnarray} 
Note that the conditions (\ref{rhon}) remove the dependence on the
scale $\mu$
 in (\ref{rholog}), thus we can drop it. No standard
requirement of
 positivity for the spectral strength, $\rho(\omega)$,
is made (see Sect. \ref{ved}). 

In the present work we do
 not intend
to determine the quark spectral function from ``first principles'',
but rather look for general consequences and implicit relations
which follow from the approach. 

 The model with the spectral
regularization (\ref{eq:spec0},\ref{rho0},\ref{rhon}), 
 supplied
with couplings obtained via the gauge technique, 
 possesses {\em
simultaneously} the following features: 
\begin{enumerate}
\item Gives finite values for hadronic observables, which can be used
to fix the 
 inverse moments, (\ref{rhoinv}), and the log moments,
(\ref{rholog}).
\item Satisfies by construction the electromagnetic and chiral Ward-Takahashi identities, 
 thus reproducing all the necessary symmetry requirements.
\item Satisfies the anomaly conditions.
\item Complies to the QCD factorization property, in the sense that
the 
 expansion of a correlator in a large-momentum $Q$ is a pure
twist-expansion involving
 only the 
 inverse powers of $Q^2$,
without the $\log Q^2$ corrections. Thus, the model can be used 
 to
compute the soft matrix elements needed to describe the deep inelastic
inclusive and exclusive processes and to analyze the pion structure
function, pion distribution amplitude, {\em etc.}
\end{enumerate}
The fact that all above features can be satisfied simultaneously 
in a chiral quark model is far from trivial \cite{Ru02}.

One of the advantages of our regularization is that all calculations
can be directly undertaken in the Minkowski space, although nothing
prevents us from working in the Euclidean space.  When dealing with
bound-state problems, the continuum Euclidean formulation encounters
practical difficulties in momentum space since going to the bound
state pole requires a continuation to the Minkowski region, and hence
a continuation of the quark propagator to the complex plane is
required. This is a problem since much of our phenomenological insight
is based on the behavior of the quark propagator in the Euclidean
region. In our approach the analytic continuation from the
Euclidean to the Minkowski space can be done in a straightforward
manner because the generalized Lehmann representation~(\ref{eq:spec0})
implies a definite analytic structure. 

In the language of the practitioners of chiral quark models such as
the Nambu--Jona-Lasinio model, the regularization introduced by the gauge technique is
very special because not only it makes the theory finite, but also
corresponds to taking the infinite cut-off limit in those observables
which do not depend on the constituent quark mass. This includes
the proper fulfillment of the anomalies. As we will see below, this is
a very rewarding aspect of the present investigation, which avoids the
artificial separation between the real and imaginary parts in the
Euclidean action, or, equivalently, the normal and abnormal parity
processes in the Minkowski space. At the same time, the study of
several processes in the high-energy limit turns out to be compatible
with factorization of amplitudes into hard and soft pieces in the
twist expansion precisely because of the regularization and the set of
conditions (\ref{rho0},\ref{rhon}) imposed upon the spectral function
$\rho(\omega)$. 
 Although such a behavior is expected in
perturbative QCD, it is very difficult to comply to it in traditional
chiral quark models. This point has been recently discussed by one of
us in Ref.~\cite{Ru02}, where it is pointed out that for the process
$\gamma^* \to \pi^0 \gamma $, which involves the transition form
factor $ F_{\gamma^* \pi^0 \gamma} (Q^2) $, there is a conflict
between the chiral anomaly normalization condition for $ F_{\gamma^*
\pi^0 \gamma} (0) =1/(4\pi f_\pi) $, and the expected QCD
factorization at large momenta, $Q^2 F_{\gamma^* \pi^0 \gamma} (Q^2) \to 2 f_\pi$. The
conflict 
 persists in any standard approach; to fulfill the anomaly
the absence of a regulator is required, but to achieve factorization an
explicit regularization must be considered. Our model is free of such
contradictions, and both conditions turn out to be
satisfied simultaneously. Actually, as we show in Sects. \ref{anomff} and \ref{pdf},
for the pion the leading twist contribution to the parton
distribution function (PDF), $ V_\pi (x) $, and the parton
distribution amplitude (PDA), $\varphi_\pi(x) $ the following
remarkable relation holds at the model's scale $Q_0$:
\begin{eqnarray}
\varphi_\pi (x, Q_0) = \frac12 V_\pi (x, Q_0 ) = 1.
\label{pda=pdf}
\end{eqnarray} 

We stress that the interpretation of results of low-energy models in
the context of the high-energy processes necessarily involves QCD
evolution to account for logarithmic perturbative radiative
corrections. The soft matrix elements computed in the model are
obtained at the {\em working energy scale of the model}, $Q_0$,
typically quite low, and have to be evolved to the scales realized
experimentally with help of the QCD evolution. Only then the
comparison to data can be made~\cite{JR80,Ja85}. In that sense, chiral
quark models provide the initial conditions for the QCD evolution. The
phenomenological success of Eq.~(\ref{pda=pdf}) {\em after evolution}
has been described elsewhere~\cite{DR95,DR02,RB02,Ru02}.

In Sect. \ref{sec:VMD} we provide a model for the moments
(\ref{rhoinv}) of the spectral function based on the
vector-meson dominance (VMD) in the pion electromagnetic form factor.
Actually, only the even and negative moments are determined by such a
method. Remarkably, the positive moments obtained by analytic
continuation automatically fulfill the spectral conditions,
Eq.~(\ref{rho0},\ref{rhon}), and the log-moments~(\ref{rholog}) can also be
determined. Interestingly, the inverse-moment problem for the VMD-inspired
model can be solved, yielding a simple function with a certain
cut-structure in the complex $\omega$-plane. As a result, the model becomes fully
explicit and some further results can be obtained. Using the insight
provided by the VMD model we work out in Sect.~\ref{sec:q_vmd} the
quark propagator, which possesses a certain cut structure but has {\em no poles in
the whole complex plane}. The quark mass function agrees remarkably
well with recent lattice data~\cite{Bowman:2002bm,Bowman:2002kn}.  We
illustrate the power of the method in Sect.~\ref{other} by presenting
some further predictions based on the VMD model, namely, the pion transition
form factor, the pion light cone wave function, the non-local quark condensate, 
and the unintegrated parton distribution function of the pion. 

\section{Quark propagator \label{sec:quark}}

\subsection{Generalized spectral representation for the quark propagator \label{qqr}} 

Our starting point is the definition of the quark propagator. In the
momentum space we have
\begin{eqnarray}
S({p}) = -i \int d^4 x e^{i p \cdot x} \langle 0 | T \left\{
q(x) \bar q(0) \right\} | 0 \rangle .
\label{eq:quark_pro} 
\end{eqnarray}
We assume the spectral representation (\ref{eq:spec0}) for the quark
propagator, where $\rho(\omega)$ is the spectral function, and $C$ is
a contour in the complex $\omega-$plane chosen in a suitable way. We
do not specify explicitly what the contour $C$ is, hence the
representation (\ref{eq:spec0}) is a  generalization of the standard
Lehmann representation \cite{IZ80}. To see the connection, let us
consider the special example of a contour $C$ running from $-\infty$
under the real negative axis and crossing through zero above the real
positive axis going to $+\infty$, yielding the form proposed in
Ref.~\cite{DW77}
\begin{eqnarray}
S({p}) = \int_{-\infty}^\infty d \omega { \rho( \omega ) \over
\slashchar{p} - \omega +{i} \epsilon (\omega) } ,
\label{eq:leh_rep1} 
\end{eqnarray}
where $ \epsilon(\omega) = 0^+ {\rm sgn} (\omega)$.  Squaring the
denominator in Eq.(\ref{eq:leh_rep1}) yields a more customary form of
the Lehmann representation~\cite{IZ80},
\begin{eqnarray}
S (p)&=& \int_0^\infty d \omega \frac{\slashchar{p} \rho_V (\omega) + \omega \rho_S
(\omega)}{p^2 - \omega^2 + {i} 0^+} ,
\label{eq:leh_rep2} 
\end{eqnarray} 
where the vector and scalar spectral functions 
\begin{eqnarray}
\rho_V (\omega) &=& \rho(\omega)+ \rho(-\omega),  \nonumber \\ 
\rho_S (\omega) &=& \rho(\omega) - \rho(-\omega),  
\label{eq:r_r1r2}
\end{eqnarray} 
are defined for $\omega \ge 0$ respectively. Notice that the functions
$\rho_V(\omega)$ and $\rho_S(\omega)$ are independent of each
other. Equivalently, the values of $\rho(\omega)$ 
 for positive and
negative values of $\omega$ are also independent of each other.  
The quark propagator in the standard Lehmann representation has an
analytic structure of poles and cuts on the real axis of the complex
$p^2 -$plane, where the positivity conditions $\rho_V(\omega) \ge 0 $
and $ \rho_V (\omega) \ge \rho_S (\omega) $ for $\omega \ge 0$
hold. With
 Eqs.~(\ref{eq:r_r1r2}), the positivity conditions are
equivalent to saying that $ \rho(\omega) \ge 0 $ for any $\omega$, positive
or negative. They follow from 
 the requirement of a physical
Hilbert. We will see in Sect. \ref{ved} that chiral symmetry breaking,
together with the finiteness of hadronic observables, implies that in
our case a {\it real} $\rho(\omega)$ cannot be positive definite,
which is a simple consequence of the conditions
(\ref{rho0},\ref{rhon}) if the contour is taken to be the standard
one. Actually, the particular realization proposed below in
Sect.~\ref{sec:VMD} based on the vector-meson dominance of the
electromagnetic pion form factor shows that one needs in fact a
non-trivial contour to avoid end-point singularities.  We refrain from
speculating on the possible connection of this fact to an indication
of quark confinement, nevertheless it is certainly true that a non-positive
spectral function cannot be understood in terms of physical particles
on the mass shell. On the other hand, a non-trivial quark propagator
must depend non-trivially on momentum if it is defined in the whole
complex plane. As will become clear in Sect. \ref{sec:q_vmd}, a 
VMD-based model produces a complex spectral function $\rho(\omega)$ on a {\it
complex} contour $C$, which results in a quark propagator with cuts only.

In a model without confinement the spectral representation of a
propagator is a well defined concept. In a gauge theory, like QED, the
spectral representation depends on the particular gauge, because the
two-point function does. In a theory with confinement not much is
known about the analytic properties of the quark propagator, except
for the fact that poles at real positive values of $p^2 $ with
positive residues are certainly excluded. In QCD such a representation
certainly exists in perturbation theory where confinement is not
manifest. The study of $\rho(\omega)$ within QCD yields at LO, for
$m \to 0$, the following expression \cite{Ha88}:
\begin{eqnarray}
\rho(\omega) = \delta (\omega-m) + {\rm sign} (\omega) \frac{\alpha C_F}{4\pi}
\frac{1-\xi}\omega \theta (\omega^2-m^2) , 
\end{eqnarray} 
where $\xi$ is the gauge parameter. However, if the general 
representation is valid, its detailed properties may be quite
different in the non-perturbative regime. 

As will become clear below, the real strength of the
ansatz~(\ref{eq:spec0}) relies in the fact that having assumed certain
properties of $\rho(w)$ reduces the calculations of physical
observables to nothing more than the standard one-loop analysis. In
addition, it allows for going from the Euclidean to the Minkowski
space, back and forth. Thus, our assumption is essentially that of
analyticity of the quark propagator $ S(p)$, and the possibility to
analytically continue it in the whole complex plane.  This looks, in
principle, very different from the approach invoked in non-local
models, where only the Euclidean region is used to justify the
propagator. Nevertheless, calculation of physical observables require
in practice an extrapolation into the complex $p^2$-plane, which by
itself can only be justified through analytic continuation. The
previous argument does not justify Eq.~(\ref{eq:spec0}), but it shows
that we are not making any additional assumptions as compared to those
implied in non-local models.

Although in perturbation theory the integration contour may be kept on the
real axis, there are cases where singularities may pinch the
integration path. This circumstance becomes a problem, since either
analyticity or relativistic invariance may be spoiled. In this regard
a number of prescriptions have been devised in order to avoid such a
situation~\cite{LW69,CL69,BG84}. Thus, in general, we will assume that
the integration path is an arbitrary contour chosen in a convenient
way. This contour integration requires effectively considering complex
masses. Genuine non-local models formulated in the Minkowski space also
require a specification of the integration contour in momentum space
in order to keep relativistic invariance~\cite{PR01}.  
An example of a definite prescription of the choice of such a contour is given in Sect. {\ref{sec:VMD}.

The quark propagator (\ref{eq:spec0}) may be parameterized in the standard form, 
\begin{equation}
S(p)=A(p)\p+B(p)=Z(p)\frac{\p +M(p)}{p^{2}-M^{2}(p)}, \label{Snl}
\end{equation}
with
\begin{eqnarray}
A(p)&=&\int d\omega \frac{\rho (\omega )}{p^{2}-\omega ^{2}},\nonumber \\
B(p)&=&\int d\omega \frac{\rho (\omega )\omega}{p^{2}-\omega ^{2}},\label{ABdef}
\end{eqnarray}
and the mass and wave function renormalization functions given by 
\begin{eqnarray}
M(p) &=&\frac{B(p)}{A(p)},  \label{Mp} \\
Z(p) &=&(p^{2}-M^{2}(p))A(p), \label{Zp}
\end{eqnarray}
respectively. Let us note that if we had $\rho(\omega)=\rho(-\omega)$,
then the quark mass would vanish, $M(p^2)=0$, and  spontaneous breaking of the
chiral symmetry would then be precluded. Thus, in general, we expect 
$\rho(\omega)$ not to be an even function. In the following sections
we will compute one-by-one the physical observables and accumulate conditions that
have to be satisfied by the moments of the  spectral function
$\rho(\omega)$. 
 
\subsection{Quark condensate \label{qq}} 

With help of the representation (\ref{eq:spec0}) the quark
condensate (for a single flavor) may be straightforwardly computed, yielding
\begin{eqnarray}
\langle \bar q q \rangle &=& - i N_c \int d\omega \rho(\omega ) \int {d^4 p \over
(2\pi)^4 } {\rm Tr} {1\over \slashchar{p}-\omega } \nonumber \\ &=& -  4 i 
N_c \int d\omega  \rho(\omega ) \int {d^4 p \over (2\pi)^4 } {\omega \over p^2-\omega ^2},
\end{eqnarray}
where the trace is over the Dirac space, and $N_c=3$ is the number of colors.
The integral over the momentum $p$ is quadratically divergent. This
requires the use of an auxiliary regularization method,
{\em removed} at the end of the calculation. With
three-dimensional cut-off, $\Lambda$, one gets for large $\Lambda$ 
\begin{eqnarray}
\langle \bar q q \rangle &=& - { N_c \over 4 \pi^2} \int d\omega  \omega 
\rho(\omega ) \left [ 2\Lambda^2 + \omega^2 \log \left ( \frac{\omega^2}{4 \Lambda^2}\right )+\omega^2
\right ]. \nonumber \\
\end{eqnarray}
The finiteness of the result at $\Lambda \to \infty $ requires the conditions
\begin{eqnarray}
\rho_1=0, \;\;\; \rho_3=0,
\end{eqnarray}   
and thus 
\begin{eqnarray}
 \langle \bar q q \rangle = - {N_c \over  4 \pi^2} 
  \int d\omega \log(\omega^2) \omega ^3  \rho(\omega ) =  -{N_c \over  4 \pi^2} \rho'_3 .
\end{eqnarray}
Exactly the same conclusions are reached if the four-momentum auxiliary regularization is introduced.
Note that the $\rho_3=0$ spectral condition allowed for rewriting $\log(\omega^2/\Lambda^2)$ as 
$\log(\omega^2)$, hence {\em no scale dependence} (no ``dimensional transmutation'') is present in the 
final expression.
The dimensional regularization in $4+\varepsilon$ dimensions,
gives
\begin{eqnarray}
\langle \overline{q}q\rangle &=&\frac{N_{c}}{4\pi ^{2}}\int d\omega 
\rho (\omega  )\omega  ^{3}\left[ -\log \left( \omega  ^{2}/\mu
^{2}\right) -\frac{1}{\varepsilon }+1\right] .  \nonumber \label{qq2}
\end{eqnarray}
However, the dimensional regularization ``hides'' some conditions, for
instance here it leads only to the $\rho_3=0$ condition, and does not
require the $\rho_1=0$ condition. This is due to the fact that in 
the dimensional regularization the
power divergences have a fixed ratio. 

Finally, we remark that in the
perturbative phase with no spontaneous symmetry breaking, where 
$\rho(\omega )=\rho(-\omega )=\delta(\omega)$, we have $\langle \bar q
q \rangle =0$.
 With the accepted value of 
\begin{eqnarray}
\langle \bar q q \rangle =
 \langle \bar u u \rangle =\langle \bar d d \rangle \simeq -(243~{\rm MeV})^3  \label{qqioffe}
\end{eqnarray} 
(at the typical hadronic scale of $0.5-1$~GeV) \cite{Io02} we can infer the value of 
the third log-moment, $\rho'_3$. 
The sign of the quark condensate shows that 
\begin{equation}
\rho'_3 > 0.
\end{equation}  

\subsection{Non-local quark condensate} \label{nlqcon}

In various QCD studies (see
e.g. Refs.~\cite{Mikhailov:1989nz,Mikhailov:1992,Bakulev:2002hk,Dorokhov:2001wx} and
references therein) the non-local quark condensate,
\begin{eqnarray}
Q(x) \equiv \frac{\langle q(0) q(x) \rangle}{\langle q(0) q(0) \rangle}, \label{qx}
\end{eqnarray}
plays an important role. In our approach 
\begin{eqnarray}
\langle \bar q(0) q(x) \rangle &=& - 4 i N_c \int d\omega \rho(\omega
) \int {d^4 p \over (2\pi)^4 } {\omega \over p^2-\omega ^2} e^{i p
\cdot x}, \nonumber \\
\end{eqnarray}
and, consequently, after performing the Fourier-Bessel transform, 
\begin{eqnarray}
Q(x) = \frac{1}{\rho'_3} \int d\omega \rho(\omega ) 4 \omega^3
\frac{K_1(\sqrt{-\omega^2 x^2})}{\sqrt{-\omega^2 x^2}},  \label{qxw}
\end{eqnarray}
where $x$ denotes the Minkowski coordinate.
A related quantity is the average vacuum virtuality of the quarks,
$\lambda_q$, defined through
\begin{eqnarray}
\lambda_q^2 \equiv \frac{\langle \bar q (\partial^2) q \rangle }{\langle \bar q q \rangle}.
\end{eqnarray}
With our spectral regularization method we find that 
\begin{eqnarray}
\frac{\langle \bar q (\partial^2)^n q \rangle }{\langle \bar q q \rangle}
&\equiv& \frac{\int d\omega  \rho(\omega ) \int {d^4 p } {(-p^2)^n\omega / (p^2-\omega ^2)}}
{\int d\omega  \rho(\omega ) \int {d^4 p } {\omega / (p^2-\omega ^2)}} \nonumber \\
&=&(-)^n \frac{\rho'_{2n+3}}{\rho'_3},
\label{eq:nlcon_mom} 
\end{eqnarray}
in particular
\begin{eqnarray}
\lambda_q^2 = -\frac{\rho'_5}{\rho'_3},
\end{eqnarray}
while the QCD sum rules estimates 
suggest the value $\lambda_q^2 \simeq 0.5 \pm 0.1~{\rm GeV}^2$ \cite{BI82}.
The positivity of $\lambda_q^2$ and $\rho'_3$ (see Sect. \ref{qq}) 
enforces 
\begin{equation}
\rho'_5 < 0.
\end{equation}

\subsection{Vacuum energy density \label{ved}} 

Continuing the quest for the conditions on the spectral function $\rho(\omega )$
we now study the vacuum energy density. 
The energy-momentum tensor for a purely quark model is defined as
\begin{eqnarray}
 \theta^{\mu \nu} (x) = \bar q (x) \frac{i}{2} \left\{ \gamma^\mu
\partial^{\nu} + \gamma^\nu \partial^{\mu} \right\} q(x)- g^{\mu \nu}
{\cal L} (x). \nonumber \\ \label{eq:en-mom} 
\end{eqnarray} 
At the one-quark-loop level 
\begin{eqnarray}
&& \langle \theta^{\mu \nu} \rangle = - i N_c N_f \int d\omega  \rho(\omega )
\int {d^4 p \over (2\pi)^4 } \times \nonumber \\ && {\rm Tr} {1\over \slashchar{p}-\omega } \left[
\frac12 \left( \gamma^\mu p^\nu + \gamma^\nu p^\mu \right)- g^{\mu
\nu} ( \p - \omega  ) \right] \nonumber \\ && = - 4 i N_c N_f \int d\omega 
\rho(\omega ) \int {d^4 p \over (2\pi)^4 } {p^\mu p^\nu - g^{\mu \nu}
(p^2-\omega^2) \over p^2-\omega^2} \nonumber \\ 
&& = B g^{\mu \nu} + \langle \theta^{\mu \nu} \rangle_0 ,
\end{eqnarray}
where $N_f$ is the number of flavors and  $ \langle \theta^{\mu \nu}
\rangle_0 $ is the energy-momentum tensor for the free theory, {\em
i.e.} evaluated with $\rho(\omega)=\delta(\omega)$.  The quantity $B$
is the vacuum energy density given by
\begin{eqnarray} 
B= - i N_c N_f 
\int d\omega  \rho(\omega ) \int {d^4 p \over (2\pi)^4 } {\omega^2 \over p^2-\omega^2}, \label{Bint}
\end{eqnarray} 
where in the subtraction of the free part we have used the spectral condition (\ref{rho0}).
The integral over $p$ is quadratically divergent, but there is an
additional power of $\omega$ as compared to the case of the quark condensate. Hence, the
conditions that have to be fulfilled for $B$ to be finite are
$\rho_2=0 $ and  $\rho_4 = 0$. Then
\begin{eqnarray}
B = - {N_c N_f \over 16 \pi^2} \rho'_4 = - {3 N_c \over 16 \pi^2} \rho'_4  \label{eq:w4log} 
\end{eqnarray}
for three flavors, $N_f=3$, used from now on. 

Interestingly, 
the even conditions (here quadratic and quartic) imply that $\rho(\omega)$ {\em cannot be
positive definite}; otherwise the even moments could not vanish. 

An alternative expression
for $B$ can be obtained from integrating by parts in the variable $p$ and
using the conditions $\rho_2=\rho_4=0$,
\begin{eqnarray} 
B =  2i N_c N_f \int d\omega  \rho(\omega ) \int {d^4 p \over (2\pi)^4 } \log(
p^2-\omega ^2 ) ,
\end{eqnarray} 
which reminds us of the vacuum energy density calculated in the
effective action formalism. Another interesting version of $B$ is obtained by
integrating Eq.~(\ref{Bint}) with respect to the variable $p_0$ first, when we get
\begin{eqnarray}
B = - 2 N_c N_f \int d\omega  \rho(\omega ) \int \frac{d^3 k}{(2\pi)^3} \sqrt{k^2+\omega ^2}.
\end{eqnarray}
The interpretation of this equation is obvious: $B$ is a weighted
integral of negative energy quarks with constituent mass
$\omega$. Upon imposing the $\rho_2=0$ and $\rho_4=0 $ conditions the
integration of the three momentum integral yields
Eq.~(\ref{eq:w4log}). 
 
In the case of spontaneous chiral symmetry breaking one expects, 
$ \langle \theta_{00} \rangle < \langle \theta_{00} \rangle_0 $, or $B < 0$. 
According to the most recent QCD sum rules analysis for charmonium ~\cite{IoZa,Io02},
one has, for three
flavors,
\begin{eqnarray}
B = -\frac9{32} \langle \frac{\alpha}\pi G^2 \rangle =
 - ( 224^{+35}_{-70} {\rm MeV} )^4 . \label{Bqcd}
\end{eqnarray} 
Note the large uncertainty in this result. 
The negative sign of $B$ enforces 
\begin{equation}
\rho'_4 > 0.
\end{equation}

\subsection{Low- and high-momentum behavior} 

At low momenta we may formally expand the quark propagator (\ref{eq:leh_rep1}) to obtain
\begin{eqnarray}
S(p) = - \sum_{k=0}^\infty \int d\omega \frac{\rho(\omega)}{\omega} 
\left ( \frac{\p}{\omega} \right )^k = - \sum_{k=0}^\infty \rho_{-k-1} \p^k. \label{SlowQ}
\end{eqnarray} 
Hence the low-momentum expansion involves the inverse moments (\ref{rhoinv}). In particular,
for $M$ and $Z$ of Eqs. (\ref{Mp},\ref{Zp}) we find 
\begin{eqnarray}
M(0) &=& \frac{\rho_{-1}}{\rho_{-2}},  \nonumber \\
\left . \frac{dM(p^2)}{dp^2} \right |_{p^2=0} 
& = & \frac{\rho_{-3}}{\rho_{-2}}-\frac{\rho_{-1}\rho_{-4}}{\rho_{-2}^2}, \nonumber \\
Z(0) &=& \frac{\rho_{-1}^2}{\rho_{-2}}.  \label{SlQ}
\end{eqnarray}
A knowledge of these quantities, for instance from lattice
calculations, 
 would help to determine the inverse moments
(\ref{rhoinv}) and 
 constrain the spectral function. According to
Eqs. (\ref{SlQ}), the positivity of $Z(0)$ leads to $\rho_{-2}>0$,
while  combined with the positivity of $M(0)$ gives $\rho_{-1}>0$.

In the limit of large momentum, $p \to \infty$, we formally have
\begin{eqnarray}
S(p) &\sim& \frac1{\slashchar{p}} \int d\omega \rho(w) +
\frac1{p^2} \int d\omega \omega  \rho(\omega ) \label{ShighQ} \\ &+& \frac1{\slashchar{p}^3 }
\int d\omega \omega^2 \rho(\omega) 
% +  \frac1{{p}^4} \int d\omega \omega^3 \rho(\omega)
 + ... \nonumber 
\end{eqnarray} 
Normalization of the quark propagator in the asymptotic region to
$1/\p$ leads to the condition $\rho_0 = 1$. Furthermore, since
$M(p^2)$ should vanish asymptotically, we conclude that $\rho_1 = 0$. 

Note also that if all spectral conditions (\ref{rho0},\ref{rhon}) are
assumed then the asymptotic large momentum expansion would yield a
trivial free massless quark propagator. Thus, the high-momentum
expansion cannot represent the full (and non-trivial) propagator. This
indicates some non-meromorphic structure in $p$ at infinity. In
Sects. \ref{sec:VMD} and \ref{sec:q_vmd} we will present a particular
realization of this situation.

\section{Gauge technique and the vertex functions \label{sec:gauge}} 

Up to now we have accumulated the lowest spectral conditions
(\ref{rho0},\ref{rhon}), up to $n=4$.  Higher conditions will appear
in the next sections by requesting the twist expansion of correlation
functions. First, however, we need to introduce the coupling of
currents to quarks.

In QCD, the vector and axial
currents are defined as
\begin{eqnarray}
J_V^{\mu,a} (x) &=& \bar q(x) \gamma^\mu {\lambda_a \over 2} q(x) ,\\
J_A^{\mu,a} (x) &=& \bar q(x) \gamma^\mu \gamma_5 {\lambda_a \over 2} q(x) .
\end{eqnarray} 
Although our formulas below are valid for the $SU(N_f)$ flavor symmetry,
in this paper we will be concerned mainly with pion properties. This will be
understood by replacing the Gell-Mann matrices, $\lambda_a$, by the
Pauli matrices, $\tau_a$. Conservation of the vector current (CVC) and
partial conservation of
 the axial current (PCAC) implies that
\begin{eqnarray}
\partial_\mu J_V^{\mu,a} (x) &=& 0, \\ \partial_\mu J_A^{\mu,a} (x) &=&
\bar q(x) \hat M_0 i \gamma_5 {\lambda_a \over 2} q(x),
\end{eqnarray} 
with $\hat M_0 = {\rm diag} (m_u, m_d , m_s)$ denoting the quark mass
matrix. Obviously, any effective theory of QCD must incorporate
these constraints. CVC and PCAC imply a set of flavor-gauge and
chiral Ward-Takahashi identities among correlation functions
involving vector currents, axial currents, and quark field
operators, which are based on the local current-field commutation
rules~\cite{Re67}, 
\begin{eqnarray}
\left[ J_V^{0,a} (x) , q(x') \right]_{ x_0 = x'_0 } &=& - \gamma_5
{\lambda_a \over 2} q(x) \delta(\vec x-\vec x'), \nonumber  \\ 
\left[ J_A^{0,a} (x) , q(x') \right]_{ x_0 = x'_0 } &=& - \gamma_5 
{\lambda_a \over 2} q(x) \delta(\vec x-\vec x') .
\end{eqnarray} 
A number of results are then obtained essentially for free. In the
low-energy regime pions arise as Goldstone bosons, and the standard
current algebra properties hold. In the high-energy regime, parton
model features such as scaling and the spin-1/2 nature of hadronic
constituents may be recovered. If one restricts to the one-quark-loop
approximation, the results also provide a particular solution to the
large-$N_c$ counting rules at the leading order.

To solve the Ward-Takahashi identities we follow the {\em gauge technique}
proposed in Ref.~\cite{DW77} (see also Ref.~\cite{De79,De99}) which has the
nice feature of linearizing the equations, since they deal with {\it
unamputated} Green functions. This is in contrast to the more standard
approach of writing the Ward-Takahashi identities 
for {\it amputated} Green functions, in which case
non-linear equations arise.  The gauge technique has been mostly used
in the past as a way to
 obtain solutions to the Schwinger-Dyson
equations, both in
 QED~\cite{DW77,Sa02} and in QCD~\cite{Ha88}. Only
recently it has been used to study hadron phenomenology~\cite{Ru01}. 

\subsection{Vertices with One Current} 

The vector and axial unamputated vertex functions are defined
as
\begin{eqnarray}
&& \Lambda_V^{\mu,a} (p' , p ) = i S(p') \Gamma_V^{\mu,a} (p' , p ) i S(p) \\
&=&  \int d^4 x d^4 x' \langle 0 | T \left\{ J_V^{\mu , a} (0)
q(x') \bar q(x) \right\} | 0 \rangle e^{i p' \cdot x'-i p \cdot x}, \nonumber \\
&&\Lambda_A^{\mu,a} (p' , p ) = i S(p') \Gamma_A^{\mu,a} (p' , p ) i S(p) \\
&=&  \int d^4 x d^4 x' \langle 0 | T \left\{ J_A^{\mu , a} (0)
q(x') \bar q(x) \right\} | 0 \rangle  e^{i p' \cdot x'-i p \cdot x}, \nonumber
\end{eqnarray}
respectively. Here the $\Gamma$'s represent the corresponding
amputated vertex functions~\footnote{Our conventions are as follows:
We take $e^{-i p \cdot x}$ for in-going and $e^{i p\cdot x}$ for
out-going particles.  For free massless particles the  irreducible
functions are normalized according to $ \Gamma_V^{\mu,a} =\gamma^\mu
\frac{\lambda_a}2 $ and $\Gamma_A^{\mu,a} =\gamma^\mu \gamma_5
\frac{\lambda_a}2 $.  The convention for the Dirac matrices and the metric
tensor is the one of Ref.~\cite{IZ80}.}. The Ward-Takahashi identity
for the full vector-quark-quark vertex reads
\begin{eqnarray}
(p'-p)_\mu \Lambda_V^{\mu,a} (p' , p ) = S(p') {\lambda_a \over 2}-
{\lambda_a \over 2} S(p) \, . \label{VWT}
\end{eqnarray} 
Likewise, for the axial-quark-quark vertex  we have 
\begin{eqnarray}
 (p'-p)_\mu \Lambda_A^\mu (p', p) = S(p') {\lambda_a \over 2} \gamma_5
 + \gamma_5 {\lambda_a \over 2} S(p) \, . 
\label{AWT}
\end{eqnarray} 

The gauge technique, introduced in Ref.~\cite{DW77}, consists of
writing a solution for the vector unamputated vertex in the form
\begin{eqnarray}
\Lambda_V^{\mu,a} (p', p) = \int d \omega \rho(\omega ) { i \over
\slashchar{p'} - \omega } \gamma^\mu {\lambda_a \over 2} { i \over
\slashchar{p} - \omega } . \label{unaV}
\end{eqnarray} 
The axial-vertex ansatz reads   
\begin{eqnarray}
\Lambda_A^\mu (p', p) &=& \int d \omega \rho(\omega ) { i \over
\slashchar{p'} - \omega } \left( \gamma^\mu - {2 \w q^\mu \over q^2}
\right) \gamma_5 {\lambda_a \over 2} { i \over \slashchar{p} - \omega
}. \nonumber \\
\label{unaA}
\end{eqnarray} 
In this way the Ward-Takahashi identities are linearized. It can be
readily verified that these ans\"atze fulfill the identities
(\ref{VWT},\ref{AWT}), respectively, up to {\em undetermined transverse
pieces}.  A consequence of the axial Ward-Takahashi identity is the
occurrence of a massless pseudoscalar pole identified with the pion,
which takes place only if $\rho(\omega) \neq \delta(\omega)$. 

The pion wave function, corresponding to the $\pi \to q \bar q $
vertex, is defined as
\begin{eqnarray}
\Lambda_\pi^a (p+q,p) = i \int d^4 x e^{- i p \cdot x} \langle 0 | T
\left\{ q(0) \bar q(x) \right\} | \pi_a (q) \rangle . \nonumber \\
\end{eqnarray} 
Near the pion pole we get 
\begin{eqnarray}
\Lambda_A^{\mu,a} (p+q,p) \to - {q^\mu \over q^2} \Lambda_\pi^a (p+q,p),
\end{eqnarray} 
where the pion wave function is given by 
\begin{eqnarray} 
\Lambda_\pi^a (p+q,p) = \int d \omega  \rho(\omega ) { i \over \slashchar{p}+
\slashchar{q} - \omega } \frac{2 \w}{f_\pi} \gamma_5 {\lambda_a \over 2} { i
\over \slashchar{p} - \omega}. \nonumber \\
\label{eq:gt} 
\end{eqnarray} 
We recognize in our formulation the Goldberger-Treiman relation for
quarks; under the spectral integral over $\omega$ the pseudoscalar
coupling of a pion to the quarks is the ratio of the spectral quark
mass to the pion weak decay constant, 
\begin{eqnarray}
g_\pi (\omega) = \frac{\omega}{f_\pi}. \label{GTrel}
\end{eqnarray} 

\subsection{Vertices with two Currents} 

The vertices with two currents, axial or vector, will be needed below
when computing form factors. We define the axial-axial vertex (other
vertices can be done in a similar fashion) by
\begin{eqnarray} 
&& (2\pi)^4 \delta^{(4)} (p'+q'-p-q) 
\Lambda_{AA}^{\mu,a;\nu,b} (p',q'; p,q ) \nonumber \\ &&=  \int d^4 x
d^4 x' d^4 y' d^4 y  e^{i ( q' \cdot x' + p' \cdot y' - q \cdot x -p
\cdot y )} \nonumber \\ && \times \langle 0 | T \left\{ J_A^{\mu , a} (x)
J_A^{\nu , b} (x') q(y) \bar q(y') \right\} | 0 \rangle , 
\end{eqnarray} 
which in SU(2) fulfills the Ward-Takahashi identity 
\begin{eqnarray}
-i q^\mu  \Lambda_{AA}^{\mu,a;\nu,b} (p',q' ; p , q ) &=&  i \epsilon_{bac} 
\Lambda_V^{\nu,c} (p',p)  \\
+  {\tau_a \over 2} \gamma_5  \Lambda_A^{\nu b} (p'-q,p) 
&+&  \Lambda_A^{\nu b} (p',p+q) {\tau_a \over 2} \gamma_5 .
\nonumber 
\end{eqnarray} 
Up to transverse pieces one gets the solution 
\begin{eqnarray}
&& \Lambda_{AA}^{\mu,a;\nu,b} (p',q'; p,q ) = \int d \w \rho(\w)
{i\over \slashchar{p}' -\w} \Big\{ \nonumber \\ && \left( \gamma^\nu
- {2 \w {q'}^\nu \over {q'}^2} \right) \gamma_5 {\tau_b \over 2}
{i\over \slashchar{p} + \slashchar{q} -\w} \left( \gamma^\mu - {2 \w
q^\mu \over q^2} \right) \gamma_5 {\tau_a \over 2} \nonumber \\ &+&
{\rm crossed} + {\w {q'}^\nu q^\mu \over {q'}^2 q^2} \delta_{ab}
\Big\} {i\over \slashchar{p} -\w} .
\end{eqnarray} 
The $\pi q \to \pi q $ unamputated scattering amplitude is defined as
\begin{eqnarray}
&& (2\pi)^4 \delta^{(4)} (p'+q'-p-q) 
\Lambda_{\pi\pi}^{ba} (p',q';p,q) \\ &=&  \int d^4 x d^4 x' e^{i ( p' 
\cdot x' -p \cdot x) } \langle \pi_b (q') | T \left\{ q(x) \bar q(x')
\right\} | \pi_a (q) \rangle . \nonumber 
\end{eqnarray} 
At the pion poles, $q^2 , {q'}^2 \to 0 $, we get
\begin{eqnarray}
\!\!\!\!\!\! \Lambda_{AA}^{\mu,a;\nu,b} (p',q'; p,q ) \to { {q'}^\nu q^\mu \over
q^2 {q'}^2 } f_\pi^2 \Lambda_{\pi\pi}^{ba} (p',q';p,q) ,
\end{eqnarray} 
where 
\begin{eqnarray}
\!\!\!\!\!\! && \Lambda_{\pi\pi}^{ba} (p',q'; p,q ) = \int d \w \rho(\w) {i\over
\slashchar{p}' -\w} \Big\{ {\w \over f^2 } \delta_{ab} \nonumber \\ \!\!\!\!\!\! && +
{\w \over f} \gamma_5 \tau_b {i\over \slashchar{p} +
\slashchar{q} -\w} { \w \over f} \gamma_5 \tau_a + {\rm
crossed} \Big\} {i\over \slashchar{p} -\w} . 
\label{eq:piq_piq}
\end{eqnarray} 

\section{Vacuum polarization \label{vacuum}} 

The vacuum polarization is obtained from the vector-vector correlation
function, which is constructed by closing the quark line in the
unamputated vector vertex (\ref{unaV}), with the result
\begin{eqnarray}
&& i \Pi_{VV}^{\mu a , \nu b } (q) = \int d^4 x e^{-i q \cdot x}
\langle 0 | T \left\{ J_V^{\mu a} (x) J_V^{\nu b} (0) \right\} | 0
\rangle \nonumber \\ &=& -N_c \int {d^4 p \over (2\pi)^4 } {\rm Tr} \left[
\Lambda_V^{\mu,a} ( p+q , p ) \gamma_\nu {\lambda_b \over 2}
\right] \nonumber \\ &=& - N_c \int d\omega \rho(\omega) \int {d^4 p \over
(2\pi)^4 } \times \nonumber \\
&& {\rm Tr} \left[ {i\over \slashchar{p}-\slashchar{q} - \omega}
\gamma_\mu {\lambda_a \over 2} {i\over \slashchar{p} - \omega} 
\gamma_\nu {\lambda_b \over 2} \right].
\end{eqnarray}
We use the dimensional regularization \footnote{The use of the
dimensional regularization guarantees gauge invariance, with no
further subtractions, hence for simplicity we use it  throughout the
paper in the applications with vector and axial currents.} and
obtain
\begin{eqnarray} 
\Pi_{VV}^{\mu a , \nu b } (q) = 
\delta_{ab} \left( -g^{\mu\nu } + { q^\mu q^\nu \over q^2} \right)
\bar \Pi_V (q^2),
\end{eqnarray} 
with 
\begin{eqnarray}
\bar \Pi_V (q^2 ) &=& \frac{N_c}{3} \int \rho(\omega) d\omega \times  \\
&& \left\{ -2 \omega^2 \bar I(q^2,\omega) + q^2 (\frac{1}{3}-I(q^2,\omega)) \right\} , \nonumber
\end{eqnarray} 
where the one-loop functions $\bar I(q^2,\omega)$ and $I(q^2,\omega)$
are introduced in App.~\ref{sec:appa}. We note that the vector
wave function renormalization,
\begin{eqnarray}
Z= \Pi' (0) = \frac{N_c}3 \int \rho(\omega) d\omega \left\{
\frac{1}{3}-I(0,\omega) \right\} ,
\end{eqnarray} 
diverges, which is the case of perturbative theories, like QED, as well. 

The one loop integral satisfies the twice-subtracted dispersion
relation (see App.~\ref{sec:appa}),
\begin{eqnarray}
\bar \Pi_V (q^2) = \frac{q^4} {\pi} \int_0^\infty \frac{dt}{t^2} \,
\frac{{\rm Im} \bar \Pi_V (t) }{t-q^2- i0^+}.
\end{eqnarray} 
This is in contrast to non-local quark models formulated in the
Euclidean space, where the dispersion relation is postulated, but
never deduced.  As a matter of fact, even in local models, such as in
those with the proper-time regularization, dispersion relations do not
hold \cite{analyt} due to the presence of essential singularities
generating non-analytic structure in the complex $q^2$-plane. 

To end this section, we compute the cross section for the reaction 
$e^+ e^- \to$~hadrons. This quantity is proportional to the imaginary
part of the vacuum charge polarization operator. Asymptotically, at
large $s$, we find
\begin{eqnarray}
\sigma( e^+ e^- \to {\rm hadrons}) & \to & 
{4 \pi  \alpha_{\rm QED}^2 \over 3 s } \left(\sum_i e^2_i \right) 
\int d\omega \rho(\omega), \nonumber \\
\end{eqnarray} 
where $e_i$ is the electric charge of the quark of species $i$.
Thus, the proper QCD asymptotic result is obtained when the spectral 
normalization condition (\ref{rho0}) is imposed.

\section{Pion weak decay \label{fpi}} 

The pion weak-decay constant, defined as 
\begin{eqnarray}
\langle 0 \left| J_A^{\mu a} (x) \right| \pi_b (q) \rangle = i f_\pi
q_\mu \delta_{a,b} e^{i q \cdot x},
\end{eqnarray}
can be computed from the axial-axial correlation function. 
We insert a complete set of eigenstates into the correlator, 
\begin{eqnarray}
-i \Pi_{AA}^{\mu a; \nu b } (q) &=& \int d^4 x e^{-i q \cdot x} \langle 0 | T \left\{ J_A^{\mu a} (x)
J_A^{\nu b} (0) \right\} | 0 \rangle \nonumber \\ &=&  
i f_\pi^2 \delta_{ab} {q^\mu q^\nu \over q^2} + \dots,
\end{eqnarray}
and recover the pion pole, with the dots indicating pieces regular in
the limit $q^2 \to 0$.  The procedure of closing the quark line in the
unamputated axial vertex (\ref{unaA}) results in
\begin{eqnarray}
&& -i \Pi_{AA}^{\mu a; \nu b } (q) = - N_c \int {d^4 k \over
(2\pi)^4 } {\rm Tr} \left[ \Lambda_A^\mu ( k+q , k ) \gamma_\nu
\gamma_5 {\lambda_b \over 2} \right] \nonumber \\
&& = -  N_c \int d\omega \rho(\omega ) \int {d^4 k \over (2\pi)^4 }
\times \label{piaa} \\
&& {\rm Tr}
\left[ {i\over \slashchar{k}-\slashchar{q} - \omega} \left( \gamma^\mu
- \frac{2 \omega q_\mu}{q^2} \right) \gamma_5 {\lambda_a \over 2} {i\over
\slashchar{k} - \omega}  \gamma^\nu \gamma_5 {\lambda_b \over 2}
\right] . \nonumber 
\end{eqnarray} 

Note that the above expression involves only one full vertex, $(
\gamma_\mu
 - 2 \omega q_\mu/q^2) \gamma_5 \lambda_a/2$, and one bare
vertex, $\gamma_\nu \gamma_5 \lambda_b/2$. This is needed to avoid
double counting, and 
 complies to the method of Pagels and Stokar
\cite{PS79}: in a two-point correlator all  diagrams of the
underlying theory (QCD) can be grouped in such a way as  to dress
the quark propagators (self-energy  renormalization), and {\em one}
of the vertices, while the other vertex remains in the form from 
the underlying theory. If both vertices were dressed and no additional
subtractions were introduced,  double counting of the diagrams of
the underlying theory would result.  

With the dimensional
regularization Eq.~(\ref{piaa}) becomes 
\begin{eqnarray} 
\Pi_{AA}^{\mu a , \nu b } (q) = \delta_{ab} \left(
-g^{\mu\nu } + { q^\mu q^\nu \over q^2} \right) \bar \Pi_A (q^2),
\end{eqnarray} 
with
\begin{eqnarray}
\bar \Pi_A (q^2) = \bar \Pi_V (q^2 ) + 4 N_c \int d\omega \omega^2 \rho(\omega) I (q^2,\omega).
\end{eqnarray}
As we can see, spontaneous breaking of chiral symmetry implies a pole in
the axial-axial correlator, with the residue proportional to the
squared pion weak decay constant. The result is 
\begin{eqnarray}
f_\pi^2 &=&  4 N_c  \int d\omega \rho(\omega ) \omega ^2 I(0,\omega). \label{fpidef1}
\end{eqnarray}
A finite value for $f_\pi$ requires the condition $\rho_2 = 0$. Then
\begin{eqnarray} 
f_\pi^2 =- {N_c \over 4\pi^2 } \int d\omega \log (\omega^2 ) \omega^2 \rho(\omega)
=-{N_c \over 4\pi^2 }\rho'_2 . 
\label{eq:w2log}
\end{eqnarray} 
Again, the spectral condition $\rho_2=0$ guarantees the absence of
dimensional transmutation for this log moment. The value of the pion
decay constant can thus be used to determine $\rho'_2$. The sign
is, obviously, 
\begin{equation}
\rho'_2 < 0.
\end{equation}

\section{Weinberg sum rules \label{Weinberg}} 

The basic idea behind the Weinberg sum rules~\cite{We67,BD75} is that at
high energies chiral symmetry breaking should be small. There are two
equivalent ways to derive expressions for these sum rules in our
model: from the absorptive parts, or from the dispersive parts of the
correlators. Both are equivalent due to the dispersion relations,
which we have shown to hold in Sect.~\ref{vacuum}.  Here we present
the derivation from the absorptive parts.  The vector-vector and
axial-axial correlation functions can be subtracted from each other,
yielding for the imaginary parts
\begin{eqnarray} 
&&\frac1\pi \left\{ {\rm Im} \bar \Pi_V (q)- {\rm Im} \bar \Pi_A (q)
\right\} \nonumber \\ &=& { N_c \over \pi^ 2 } \int d\omega \omega^2
\rho(\omega) \sqrt{1-{4\omega^2 \over q^2}} \theta(q^2 - 4 \omega^2) .
\end{eqnarray}
Next, we integrate with respect to $q^2$, and use the spectral
conditions and the definition of $f_\pi$ to get
\begin{eqnarray}
&& { 1\over \pi} \int_0^\infty { dq^2 \over q^2} \left\{ {\rm Im} \bar
 \Pi_V (q)- {\rm Im} \bar \Pi_A (q) \right\} \\ &=& {N_c \over 2 \pi^2
 } \lim_{\Lambda \to \infty} \int d\omega \omega^2 \rho(\omega)
 \left\{ \log{ \Lambda^2 \over \omega^2} - 2 \right\} = f_\pi^2,
 \nonumber
\end{eqnarray} 
which coincides with the first Weinberg sum rule. Now, if we compute 
the left-hand side of the second Weinberg sum rule, we get
\begin{eqnarray} 
&& {1\over \pi} \int_0^\infty dq^2 \left\{ {\rm Im} \bar \Pi_V (q)-
 {\rm Im} \bar \Pi_A (q) \right\} \nonumber \\ &=& {N_c \over 2 \pi^2 } \int
 d\omega \omega^4 \rho(\omega) \log \omega^2 = - \frac{8}{3} B.
\end{eqnarray} 
The result, according to the second Weinberg sum rule, should involve
on the right-hand side the quantity $m \langle {\bar q} q \rangle =
f_\pi^2 m_\pi^2$, which vanishes in the chiral limit. Instead, our
formula involves the vacuum energy density, $B$, which does not vanish.
This violation of the second Weinberg sum rule is similar to findings
in other chiral quark models, and reflects, in this regard, a
deficiency of those models as well as of the present approach (see, {\em e.g.}, 
the discussion in Ref.~\cite{Broniowski:1999dm}).  A study, to be
presented elsewhere, reveals that this should not be considered a
drawback of the spectral representation method, but rather a feature
of the  particular solution of the axial Ward-Takahashi identity
\footnote{For instance, a modification of the vector vertex by
providing additional tensor coupling is capable of curing the problem
of the second Weinberg sum rule.  This important issue will be studied
elsewhere}.

\section{Pion electromagnetic form factor \label{pionff}} 

\subsection{Form factor}

The electromagnetic form factor for a positively charged pion, $\pi^+ = u \bar d $,
is defined as
\begin{eqnarray}
\langle \pi^+ (p') | J_\mu^{\rm em} (0) | \pi^+ (p) \rangle &=& e
\Gamma_\mu^{\rm em } (p' , p)  \\ &=& (p^\mu + {p'}^\mu) e F_\pi^{\rm em}
(q^2),  \nonumber
\end{eqnarray} 
with $q=p'-p$.
Following the method of Sect. \ref{sec:gauge}, we compute the form
factor by using the $\pi q \to \pi q $ scattering amplitude, closing the
fermion line, and tracing with an electromagnetic vertex:
\begin{eqnarray}
\Gamma_\mu^{\rm em} (p',p) &=&  -N_c \int {d^4
k \over (2\pi)^4 } {\rm Tr} \left[\Lambda_{\pi\pi}^{ba} (k+q,p';k,p )
\hat Q \gamma_\mu \right]. \nonumber \\
\end{eqnarray} 
Through the use of Eq.~\ref{eq:piq_piq} we get
\begin{eqnarray}
&& \Gamma_\mu^{\rm em } (p' , p) = -N_c \int d\omega \rho(\omega ) \left( { \sqrt{2}
\omega \over f_\pi} \right)^2 \int {d^4 k \over (2\pi)^4 } \times \nonumber \\
&& {\rm Tr} \left[
\gamma_\mu {i\over \slashchar{k} - \omega } \gamma_5 {i\over
\slashchar{p}+\slashchar{k} - \omega } \gamma_5 {i\over
\slashchar{q}-\slashchar{k} - \omega } \right] .
\end{eqnarray} 
For on-shell massless pions the electromagnetic form factor reads
\begin{eqnarray}
F_\pi^{em} (q^2 ) = \frac{4N_c }{f_\pi^2} \int dw \rho(\omega ) \omega
^2 I(q^2 , \omega ) .
\label{eq:ffpi}
\end{eqnarray}
which due to Eq.~(\ref{fpidef1}) is obviously normalized to unity at $q^2=0$, 
$ F_\pi^{em} (0 )=1$. 
With help of App. \ref{sec:appa} we derive the low-momentum expansion,
\begin{eqnarray}
F_\pi^{em} (q^2 ) = 1+\frac{1}{4\pi^2 f_\pi^2} \left ( \frac{q^2 \rho_0}{6}+ \frac{q^4 \rho_{-2}}{60}+
\frac{q^6 \rho_{-4}}{240}+ \dots \right ). \nonumber \\ \label{Fexp} 
\end{eqnarray}
The mean square radius reads
\begin{eqnarray}
\langle r^2_\pi \rangle = 6 {d F \over dq^2 } \Big|_{q^2 =0} =
{N_c \over 4 \pi^2 f_\pi^2} \int d\omega  \rho(\omega )={N_c \over 4 \pi^2 f_\pi^2} , \nonumber \\
\end{eqnarray} 
which coincides with the unregularized-quark-loop
result~\cite{Tarrach:1979ta} and also shows that in the present
framework the pion is an extended object. The numerical value is
\begin{eqnarray} 
\left . \langle r^2 \rangle_\pi^{\rm em} \right|_{\rm th} = 0.34{\rm fm}^2, \qquad
\left . \langle r^2 \rangle_\pi^{\rm em} \right|_{\rm exp} = 0.44 {\rm fm}^2,
\end{eqnarray} 
which is a reasonable agreement. One should not expect a perfect
agreement since from the chiral perturbation theory it is well known
that pion-loop corrections provide a sizeable enhancement for $\langle r^2
\rangle_\pi^{\rm em}$. We note that the knowledge of the pion
electromagnetic form factor allows to determine the even negative
moments of the spectral function, {\em cf.} Eq.~(\ref{Fexp}). This
will be used in Sect. \ref{sec:VMD} to build a vector-meson dominance
model.  Based on the properties of the one loop integral $I(q^2,w)$
(see App. \ref{sec:appa}), the pion form factor also fulfills a
dispersion relation in our formalism
\begin{eqnarray}
F_{\pi}^{\rm em} (q^2) = 1 + \frac{q^2} {\pi} \int_0^\infty
\frac{dt}{t} \, \frac{{\rm Im} F_{\pi}^{\rm em} (t) }{t-q^2- i0^+}.
\end{eqnarray} 
%which, again, cannot be verified in non-local models. 

\subsection{Twist expansion and spectral conditions}

In the limit of large momentum we find, according to Eq.~(\ref{Ihighq}),
\begin{eqnarray}
&& F_\pi^{em} (q^2 ) \sim  \frac{N_c}{4 \pi^2 f_\pi^2} 
\int d \omega \rho(\omega ) \omega ^2 
\{ 2-\frac{1}{\epsilon}-\log (q^2) + \nonumber \\
&& \frac{ 2 \omega^2}{q^2} \left[ \log (-q^2/\omega^2) + 1 \right] + 
\frac{ 2 \omega^4}{q^4} 
\left[ \log (-q^2/\omega^2) - \frac{1}{2} \right] \dots \}
\nonumber \\ \label{Ftw1}
\end{eqnarray}
With help of the spectral conditions (\ref{rhon}) for $n=2,4,6,...$ we 
can rewrite this expansion in the form
\begin{eqnarray}
F_\pi^{em} (q^2 ) &\sim&  - \frac{N_c}{4 \pi^2 f_\pi^2} \left [
\frac{2\rho'_4}{q^2} + \frac{2\rho'_6}{q^4} + 
\frac{4\rho'_8}{q^6}+...   \right ] \nonumber \\
&&= -\frac{8B}{3f_\pi^2 Q^2} - \frac{N_c}{4 \pi^2 f_\pi^2} \left [
\frac{2\rho'_6}{Q^4} - 
\frac{4\rho'_8}{Q^6}+...   \right ] , 
\nonumber \\ \label{Ftw}
\end{eqnarray} 
with $Q^2=-q^2$. Note the very interesting feature: the imposition of
the spectral conditions (\ref{rhon}) removed {\it all} the logarithms
of $q^2$ from the  expansion (\ref{Ftw1}), leaving a pure expansion
in inverse powers of $q^2$. Thus, {\em factorization has been
 achieved},
which in our opinion is one of the major 
 successes of the present
approach. Conversely, in order to obtain factorization, the
conditions (\ref{rhon}) must be assumed. In the present calculation
only even moments of the spectral function $\rho(\omega)$
appeared. To involve the odd moments one needs to consider a
different quantity, for instance the scalar pion form factor. We
recall that the odd spectral conditions were also needed in
Sect.~\ref{nlqcon}. 

The leading-twist coefficient in expansion (\ref{Ftw}) has a very
simple physical interpretation: it involves the ratio of the vacuum
energy density, $B$, and $f_\pi^2$.  Finally, we remark that the pure
power behavior of Eq.~(\ref{Ftw}) is characteristic of a bound-state
object, and was obtained in non-local models~\cite{PS79} and more
recently in the instanton models~\cite{Faccioli:2002jd} and the 
Nambu--Jona-Lasinio  model~\cite{Ru02}.

It is worth stressing that {\it all} spectral conditions (\ref{rhon})
are needed. If we just impose a finite number of them, say up to order
$N$, then there appear logarithmic corrections starting at order
$N+2$, of the form $ \log (Q^2) / Q^{2 N+2}$. This is what happens in
the Nambu--Jona-Lasinio model when a Pauli-Villars regularization with
quadratic subtractions is used; the pion form factor has proper
leading twist behavior but a logarithmic contribution at subleading
twist~\cite{Ru02}.

Plugging the numbers for $B$ and $f_\pi$ we get for the leading twist
contribution
\begin{eqnarray}
Q^2 F_{\pi}^{\rm em} (Q^2) \Big|_{\rm twist-2} = && -\frac{8 B
}{3f_\pi^2 } = \left( 0.78 \pm 0.61 \right) {\rm GeV}^2,  \nonumber \\ \label{fasy}
\end{eqnarray}
where the uncertainty in the model value comes from the uncertainty in
$B$. The experimental result for the {\it full} form factor is $ Q^2
F_\pi^{\rm em} (Q^2) = 0.38 \pm 0.04 {\rm GeV}^2 $ as taken
averaging some old \cite{Bebek} and recent \cite{Volmer:2000ek} data
as compiled in Ref.~\cite{Blok} in the region $ 2 {\rm GeV}^2 < Q^2 <
6 {\rm GeV}^2 $. On the other hand, a remarkable finding from the data
is that a vector meson dominance monopole model obeying the pion
charge radius up to $ Q^2 \sim 1.6 {\rm GeV}^2 $
\cite{Amendolia:1986wj},
\begin{eqnarray}
 F_{\pi}^{\rm em} (Q^2) = \frac{\Lambda^2}{Q^2 + \Lambda^2} 
\end{eqnarray} 
This yields $ Q^2 F_{\pi}^{\rm em} (Q^2) = 0.41-0.45~{\rm GeV}^2 $
depending on whether one takes $\Lambda = ( 6 / \langle r^2 \rangle )
^{1/2} = 0.73~{\rm GeV} $ or $\Lambda=M_V =0.77~{\rm GeV} $,
respectively. This corresponds to $ Q^2 F_{\pi}^{\rm em} (Q^2)
\Big|_{\rm twist-2} = \Lambda^2 = 0.53-0.59~{\rm GeV}^2$ for the same
range of $\Lambda$ values, to be compared with our estimate,
Eq.~(\ref{fasy}). Motivated by the VMD success in describing the pion
form factor in the space-like region, we will study further
consequences of this scheme in Sec. \ref{sec:VMD}.

\section{Anomalous form factor \label{anomff}} 

\subsection{Vertex function} 

The AVV-vertex is defined as
\begin{eqnarray}
&& (2\pi)^4 \delta (p-q_1+q_2) \Lambda_{AVV}^{\mu,c; \alpha,a;
\beta,b}(p,q_1,q_2) \nonumber \\ && = i \int d^4 x_1 d^4 x_2 d^4 x \times \\
&& \langle 0 | T \left\{ J_A^{\mu , c} (x) J_V^{\alpha , a} (x_1)
J_V^{\beta,b} (x_2) \right\} | 0 \rangle e^{i ( p \cdot x - q_1 \cdot x_1 +
q_2 \cdot x_2 )}, \nonumber 
\end{eqnarray} 
where $q_1$ is ingoing, while $p$ and $q_2$ are outgoing. The solution
fulfilling the relevant Ward-Takahashi identities can be written. Going to the
pion pole, $p^2 \to 0$, yields
\begin{eqnarray} 
\Lambda_{AVV}^{\mu,c; \alpha,a; \beta,b}(p,q_1,q_2) \to {p^\mu \over
p^2} \Gamma_{\pi VV }^{c;\alpha,a;\beta,b} (p,q_1,q_2).
\end{eqnarray} 
For a neutral pion, $\pi^0$, and two photons one gets 
\begin{eqnarray}
&& \Gamma^{\mu \nu}_{\pi^0  \gamma \gamma } (q_1,q_2) = -N_c \int d \w
\rho(w) \int {d^4 k \over (2\pi)^4} \times \nonumber \\ 
&&  {\rm Tr} \left[ -{\w \over f_\pi}
\gamma_5 \tau_0 {i\over \slashchar{k}-\slashchar{q}_2-\w }i \hat Q
\gamma^\mu {i\over \slashchar{k}-\w} i \hat Q \gamma^\nu 
{i \over \slashchar{k}-\slashchar{q}_1 -\w } \right] \nonumber \\
&& + {\rm crossed},
\end{eqnarray} 
where $\hat Q= B/2 + I_3 = 1/2N_c + \tau_3 / 2 $ is the quark charge
operator. Straightforward calculation of the traces yields
\begin{eqnarray}
\Gamma^{\mu \nu}_{\pi^0  \gamma \gamma } (p,q_1,q_2)  &=& 
\epsilon_{\mu\nu \alpha \beta}q_1^\alpha p^\beta F_{\pi \gamma \gamma}
(p,q_1,q_2),
\end{eqnarray} 
where the pion transition form factor,
\begin{eqnarray}
\!\!\!\!\! F_{\pi \gamma \gamma} (p,q_1,q_2) = -\frac{8}{f_\pi} \int
d\w \rho(\w) \w^2 K(p^2,q_1^2,q_2^2,\omega),
\end{eqnarray} 
has been introduced, and the three-point loop function,
$K$, is presented in App. \ref{sec:appa}.

\subsection{Neutral pion decay} 

The amplitude for the neutral pion decay, $ \pi^0 ( p ) \to \gamma (
q_1 , \mu ) + \gamma (q_2 , \nu) $, can be directly computed from the
former expression by 
taking the on-shell photons, $q_1^2 = q_2^2 = 0$, and the
soft pion condition,  $p^2=0$. We find
\begin{eqnarray}
F_{\pi \gamma \gamma} (0,0,0) &=& -\frac{8}{f_\pi} \int d\omega
\rho(\omega) \omega^2 \frac1 i \int {d^4 k \over (2\pi)^4}
\frac1{(k^2-\omega^2)^3} \nonumber \\ &=& \frac{1}{4\pi^2 f_\pi} \int
d\omega \rho(\omega) =\frac{1}{4\pi^2 f_\pi},
\end{eqnarray} 
which, when the spectral condition (\ref{rho0}) is used, coincides 
with the standard result expected from the QCD chiral
anomaly.

\subsection{Transition form factor} 

For two off-shell photons with momenta $q_1$ and $q_2$ it is
convenient to  define the photon asymmetry, $A$, and the total
virtuality, $Q^2$,
\begin{eqnarray}
A &=& \frac{q_1^2 - q_2^2}{q_1^2+ q_2^2}, \qquad -1 \le A  \le 1 \\ 
Q^2 &=& -(q_1^2 + q_2^2 ) \nonumber 
\end{eqnarray} 
or, equivalently, 
\begin{eqnarray} 
q_1^2 = -\frac{(1+A)}2 Q^2, \qquad q_2^2=-\frac{(1-A)}2 Q^2 .
\end{eqnarray} 
At the soft pion point we find
\begin{eqnarray}
F_{\pi \gamma* \gamma*} (Q^2,A) &=& -\frac{8}{f_\pi} \int dw \rho(w)
w^2 \times \label{eq:tffpi} \\ && K(0,-\frac{(1+A)}2 Q^2,-\frac{(1-A)}2 Q^2
,\omega). \nonumber 
\end{eqnarray} 
Through the use of expansion (\ref{newhq}) and the spectral conditions (\ref{rhon})
for $n=2,4,6,...$ we may write, after straightforward manipulations, 
\begin{eqnarray}
&& F_{\pi \gamma* \gamma*} (Q^2,A) = - \frac{1}{2\pi^2 f_\pi} \int_{0}^{1}dx \times \\
&& \left[ \frac{2 \rho'_2}{Q^{2}(1-A^{2}(2x-1)^{2})}- \frac{%
8\rho'_4(1+A^{2}(2x-1)^{2})}{Q^{4}(1-A^{2}(2x-1)^{2})}+...\right] . 
\nonumber
\end{eqnarray}
We can now confront this expression to the standard twist
decomposition of the pion transition form factor \cite{BL80},
\begin{eqnarray}
\!\!\! F_{\gamma^* \gamma^* \pi} (Q^2, A ) =  J^{(2)} (A)
\frac1{Q^2} +  J^{(4)} (A) \frac1{Q^4} + \dots , 
\end{eqnarray} 
which via Eqs. (\ref{eq:w2log},\ref{eq:w4log}) yields 
\begin{eqnarray} 
J^{(2)}(A) &=& \frac{4 f_\pi}{N_c } \int_0^1 dx
 \frac{\varphi_\pi^{(2)}(x)}{1-(2x-1)^2 A^2}, \label{J2} \\ 
J^{(4)}(A) &=& \frac{8 f_\pi \Delta^2 }{N_c} \int_0^1 dx \frac{
 \varphi_\pi^{(4)} (x) [1+(2x-1)^2 A^2]}
 {\left[1-(2x-1)^2 A^2\right]^2}, \nonumber \\ \label{J4}
\end{eqnarray}
with 
\begin{eqnarray}
\Delta^2 = -\frac{8B}{3 f_\pi^2}.
\end{eqnarray}
Note that this is exactly the same combination of $B$ and $f_\pi$ as in
Eq.~(\ref{fasy}). Numerically, we get 
\begin{eqnarray}
\Delta^2=(0.78 \pm 0.61)~{\rm GeV}^2.
\label{Delest}
\end{eqnarray}
An estimate made in a non-local quark model of
Ref.~\cite{Do02,Do02b} provides $\Delta^2=0.29~{\rm GeV}^2$. 
 
The form of the expansion coefficients (\ref{J2},\ref{J4}) shows that the
twist-2 and twist-4 distribution amplitudes for the pion are, at the
model working scale $Q_0$, constant and equal to unity. 
 
It is interesting to look at higher-order twist coefficients.  With
help of App.~\ref{sec:appa} and with the spectral conditions (\ref{rhon})
we get
\begin{eqnarray}
F(Q^2,A)&=&-\frac{1}{\pi^2 f_\pi} \int_0^1 du \sum_{n=0}^\infty 
\frac{\sqrt{\pi} 8^n \rho'_{2n+2}}{n! \Gamma(1/2 -n)} \times \nonumber \\
&& \left ( \frac{1}{1+A(2u-1)} \frac{1}{Q^2}\right )^{n+1} .
\end{eqnarray}
This yields the result
\begin{eqnarray}
\varphi_\pi^{(n)} (x) &=& \theta(x)\theta(1-x) \qquad {\rm for} \qquad
n=2,4,6,\dots  \nonumber \\ \label{phin}
\end{eqnarray}
All these amplitudes are by convention normalized to unity.
The prediction of the model is that they do not depend on the Bjorken $x$
variable at the scale $Q_0$.

In the limit $q_1^2 = q_2^2$, or $A=0$, we get 
\begin{eqnarray}
F_{\gamma^* \gamma^* \pi} (Q^2, 0 ) =  \frac{4 f_\pi }{N_c Q^2} \left[
1 + \frac{2 \Delta^2}{Q^2} + \dots \right] .
\end{eqnarray} 

An analogous calculation to the one presented in Refs.~\cite{RB02,Ru02} 
produces the following light-cone pion wave function in the present model:
\begin{eqnarray}
\Psi(x,k_\perp)=\frac{N_c}{4 \pi^3 f_\pi^2} \int d\omega \rho(\omega) 
\frac{\omega^2}{k_\perp^2+\omega^2} \theta(x)\theta(1-x). \nonumber \\
 \label{pionlcwf}
\end{eqnarray}
Again, this form corresponds to the low-energy scale of the model.
The pion light-cone wave function (\ref{pionlcwf})
satisfies at $k_\perp=0$ the following condition,  
\begin{eqnarray}
\Psi(x,0)=\frac{N_c}{\pi f_\pi}F_{\pi \gamma
\gamma}(0,0,0)=\frac{N_c}{4 \pi^3 f_\pi^2} \label{consist}.
\end{eqnarray}
In QCD one has a similar relation holding for quantities integrated over $x$ \cite{BL80}. In
our model this is inessential due to the fact that the $x$-dependence
is constant. This triple identity, although comes out easily in our
model, is difficult to get in local chiral models since there is a
conflict between recovering the proper anomaly and obtaining
factorization of the form factor at high photon virtualities (see a
detailed discussion in Ref.~\cite{Ru02}).

\subsection{QCD evolution} 

The results of the previous section referred to the soft energy
scale of the model. In order to compare to experimental results, obtained at large
scales, the QCD evolution must be performed. The procedure has been
discussed in detail in Ref.~\cite{RB02}, hence here we only sketch the
method and mention, for completeness, the most important
 outcomes.
For the twist-2 pion distribution amplitude the leading-order QCD
evolution is made in terms of the Gegenbauer polynomials. One begins by
interpreting our result as the initial condition,
\begin{eqnarray}
\varphi^{(2)} (x,Q_0) = \theta(x) \theta(1-x) .
\end{eqnarray}
Then the evolved distribution amplitude reads~\cite{Mu95}
\begin{eqnarray}
\varphi^{(2)} (x,Q^2) &=& 6x(1-x)\sum_{n=0}^\infty C_n^{3/2} ( 2x -1) a_n (Q), \nonumber \\
\label{ev1}
\end{eqnarray}
with
\begin{eqnarray}
a_n (Q)&=& \frac23 \frac{2n+3}{(n+1)(n+2)} \left(
 \frac{\alpha(Q^2)}{\alpha(Q_0^2) } \right)^{\gamma_n^{(0)} /(2\beta_0)} \nonumber \\
&\times& \int_0^1 dx C_n^{3/2} ( 2 x -1) \varphi^{(2)} (x ,Q_0^2), \label{ev2}
\end{eqnarray}
where $C_n^{3/2}$ are the Gegenbauer polynomials, and 
\begin{eqnarray}
\gamma_n^{(0)} &=& -\frac{8}{3} \left[ 3 + \frac{2}{(n+1)(n+2)}- 4
\sum_{k=1}^{n+1} \frac1k \right], \nonumber \\
\beta_0&=&\frac{11}{3}N_c-\frac{2}{3}N_f=9. \label{gambe}
\end{eqnarray}
With our initial amplitude we immediately get 
\begin{eqnarray}
\int_0^1 dx C_n^{3/2} ( 2 x -1) \varphi^{(2)} (x ,Q_0^2) =1 . \label{ev3}
\end{eqnarray} 
What actually matters in this analysis is the evolution ratio
$\alpha(Q^2) / \alpha (Q_0^2 ) $.  With help of
Eq.~(\ref{ev1},\ref{ev2},\ref{ev3}) we may compute the distribution
amplitude for any value of $Q^2$. The result extracted in
Ref.~\cite{SY00} and confirmed in Ref.~\cite{Bakulev:2002uc} from
experimental data~\cite{CLEO98} provides $a_2 (2.4{\rm GeV} ) = 0.12
\pm 0.03 $, hence we can fix the evolution ratio to the value
\begin{eqnarray}
\alpha(Q=2.4 {\rm GeV}) / \alpha(Q_0) = 0.15 \pm 0.06 ,
\end{eqnarray} 
which reproduces $a_2$ obtained in our model.
Then we can predict 
\begin{eqnarray}
a_4 (2.4 {\rm GeV}) &=& 0.06 \pm 0.02 \;\; ({\rm exp: } -0.14
\pm 0.03 \mp 0.09 ) , \nonumber \\ 
a_6 (2.4 {\rm GeV}) &=& 0.02 \pm 0.01 .
\end{eqnarray} 
The overall picture at the leading-twist and with leading-order QCD 
evolution is very encouraging. For further details the reader is referred to Ref.~\cite{RB02}.

\section{The $\gamma \rightarrow \pi ^{+}\pi ^{0}\pi ^{-}$ decay}

In this section we consider an example of a 
low-energy process involving a quark box diagram.
The amplitude for the decay of the photon of momentum $q$ and polarization $e
$ into three pions of momenta $p_{i}$, $\gamma (q,e)\rightarrow \pi
^{+}(p_{1})\pi ^{0}(p_{2})\pi ^{-}(p_{3})$, is equal to 
\begin{eqnarray}
&& T_{\gamma (q,\varepsilon )\rightarrow \pi ^{+}(p_{1})\pi ^{0}(p_{2})\pi
^{-}(p_{3})} =6i\int \frac{d^{4}k}{\left( 2\pi \right) ^{4}}\int d\omega
\rho (\omega ) \times \nonumber \\ && \mathrm{Tr}\left[ i\gamma _{\mu }e^{\mu }\left( \frac{1}{2N_{c}%
}\right) {i \over \slashchar{k} -\slashchar{p}_1 -\slashchar{p}_2 -\slashchar{p}_3 -w } 
\left ( - \frac{\omega}{f_\pi}\tau^+ \right ) \right.  \times 
\nonumber \\
&&\left. {i \over \slashchar{k} -\slashchar{p}_1 -\slashchar{p}_2 -w } 
\left ( - \frac{\omega}{f_\pi}\tau^0 \right ) 
{i \over \slashchar{k} -\slashchar{p}_1 -w } 
\left ( - \frac{\omega}{f_\pi}\tau^- \right ) \right. \nonumber \\ 
&& \left . {i \over \slashchar{k} -w } 
\right] \equiv F(p_{1},p_{2},p_{3})\varepsilon _{\alpha \beta \sigma \tau
}e^{\alpha }p_{1}^{\beta }p_{2}^{\sigma }p_{3}^{\tau }.
\end{eqnarray}
In the limit of all momenta going to zero we get, with the condition (\ref{rho0}), 
\begin{equation}
F(0,0,0)=\frac{1}{4\pi ^{2}f_{\pi }^{3}} \int d\omega \rho(\omega)=\frac{1}{4\pi ^{2}f_{\pi }^{3}} ,
\end{equation}
which is the correct result \cite{adler,terentev,aviv}. 

\section{Pion structure function} \label{pdf} 

As we have said in the Introduction, one of the advantages of our model over other
formulations is that calculations can be undertaken both in Minkowski
and Euclidean space. This proves crucial in the calculation of the pion structure function. 
We recall here that a Euclidean formulation allows only
for the calculation of a finite number of moments of the structure
function, $\langle x^n \rangle$, for integer $n$, requiring a
subsequent reconstruction of the distribution function. This is a
cumbersome situation. Experimental data are directly obtained in the 
$x$-space, but structure functions are difficult to pin down for large
values of $x$ (typically $x > 0.65 $). That means systematic
uncertainties for higher-order moments. 

Our following calculation also illustrates an interesting point. In the
Bjorken limit it is assumed that integrals are convergent fast enough
to allow to convert the forward Compton amplitude to a quark-target
scattering amplitude~\cite{Ja85}. We note here that while the former
corresponds to a closed quark line, the latter refers to a quark
propagator, i.e. an open quark line. In local models, such as the Nambu--Jona-Lasinio
model, the difference becomes subtle (see, {\em e.g.}, the discussion in
Ref.~\cite{Ru02}) because it is not obvious how to regulate open
quark lines. As we show below, a rewarding feature of the present
approach is that the connection from the forward Compton scattering
amplitude to the quark-target scattering formula prevails, due to the
the spectral regularization of the vertex functions. As a result, the
relation between gauge invariance and proper normalization of the PDF
remains valid.

\subsection{Derivation from the forward Compton amplitude} 

The hadronic tensor for inclusive electroproduction on the pion
reads 
\begin{eqnarray}
&& W_{\mu\nu}(p,q) = {1\over 2\pi} {\rm Im} T_{\mu\nu} (p,q) \nonumber
\\ && = W_1 ( q^2 , p \cdot q ) \left( - g_{\mu\nu} + {q_\mu q_\nu
\over q^2 }\right) \\ 
&& + {W_2 ( q^2 , p\cdot q )\over m_P^2 } \left( p_\mu -
{p \cdot q \over q^2 } q_\mu \right) \left( p_\nu - {p \cdot q \over q^2 }
q_\nu \right), \nonumber 
\end{eqnarray} 
where the forward virtual Compton scattering amplitude on the pion
is defined as 
\begin{eqnarray} 
T_{\mu\nu} ( p,q) = i \int d^4 x e^{i q \cdot x } 
\langle \pi(p)|T \Bigr\{ J_\mu^{\rm
em} (x) J_\nu^{\rm em} (0)\Bigr\}|\pi(p) \rangle . \nonumber \\
\end{eqnarray}
The amplitude can be obtained by taking the residue of the double pion
pole in the $AV \to AV$ amplitude. The gauge invariance requires
considering not only the box-like diagrams but also the process $\pi
\gamma \to \pi \to \pi \gamma$. This process may be relevant at low
energies, however it does not contribute at high energies, since it
provides higher twist contributions. The hand-bag diagrams yield
\begin{eqnarray} 
&& i T_{\mu \nu} (p,q) = - N_c \int d\omega \rho(\omega) 
\left ( {\omega\over f_\pi} \right)^2 \int {d^4 k \over
(2\pi)^2 } \times \nonumber \\ && \!\!\!\! {\rm Tr} \left[ \gamma_5 \tau_a
{1\over \slashchar{k}-\omega} \hat Q \gamma_\mu {1\over
\slashchar{k}+\slashchar{q}-\omega} \hat Q \gamma_\nu {1\over \slashchar{k}-\omega}
\gamma_5 \tau_b {1\over \slashchar{p}-\slashchar{q}-\omega} \right] \nonumber \\
&& + {\rm crossed}. 
\end{eqnarray} 
In the Bjorken limit we can make the customary approximation,
\begin{eqnarray}
\gamma^\mu {1\over \slashchar{k}+\slashchar{q}-w } \gamma^\nu &\to& 
\gamma^\mu {1\over  \slashchar{q}} \gamma^\nu \\ &=& \frac{q^\alpha}{q^2}
\left[ S^{\mu \nu \alpha \beta} \gamma_\beta + i
\epsilon^{\mu\nu\alpha\beta} \gamma_\beta \gamma_5 \right]. \nonumber
\end{eqnarray} 
The hadronic tensor is obtained as the imaginary part in the $(p+q)^2$
channel.  The Cutkosky rules amount to making the replacements 
\begin{eqnarray}
{1\over p^2 - \omega^2 } \to (-2\pi i) \theta(p_0) \delta(p^2 - \omega^2 )
\equiv (-2\pi i) \delta_+ (p,\omega), \nonumber \\
\end{eqnarray}
hence
\begin{eqnarray} 
&& W^{\mu \nu} = - N_c \int d\omega \rho(\omega) \left
( {\omega\over f_\pi} \right)^2 \times \\ 
&& \int {d^4 k \over (2\pi)^2 } (-2\pi i)^2
\frac{\delta^+ (p-k,\omega) \delta^+ (q+k,\omega)}{(k^2-\omega^2)^2 } \times \nonumber \\ &&  {\rm Tr}
\left[ \gamma_5 \tau_a (\slashchar{k}+\omega) \hat Q \gamma^\mu
( \slashchar{k}+\slashchar{q}+\omega) \hat Q \gamma^\nu (\slashchar{k}+\omega) \times  \right . \nonumber \\
&& \left . \gamma_5 \tau_b ( \slashchar{p}-\slashchar{q}-\omega) \right] . \nonumber 
\end{eqnarray}
The calculation of the traces is straightforward and the Bjorken limit
of the discontinuity can be found in App.~\ref{sec:appb}. The result
is rather simple,
\begin{eqnarray}
&& W_{\mu\nu}(p,q) = {1\over 2\pi} {\rm Im} T_{\mu\nu} ( p,q ) \to \\ 
&& F (x) \left[ - g_{\mu\nu} + {q_\mu q_\nu \over q^2 } - {1\over q^2}
( p_\mu - {q_\mu \over 2x} ) ( p_\nu - {q_\nu \over 2x} ) \right], \nonumber
\end{eqnarray}
with
\begin{eqnarray}
F (x) = {1\over 2} \sum_{i=u,d,s} e_i^2 [ \bar q_i (x) + q_i (x) ].
\end{eqnarray}
We take $\pi^+$ for definiteness and get, 
\begin{eqnarray}
u_\pi (x) = \bar d_\pi (1-x)  = \theta(x)\theta(1-x) ,
\end{eqnarray}
independent of the spectral function $\rho(\omega)$. Thus we have
recovered scaling in the Bjorken limit, the Callan-Gross relation, the
proper support, and the correct normalization. This is the same result as found
by one of us \cite{Ru01} when computing the structure function 
from the forward quark-pion
scattering amplitude, Eq.~(\ref{eq:piq_piq}), in the light-cone coordinates.  The
result has also been obtained previously by several means within the
Nambu--Jona-Lasinio model \cite{DR95,WRG99}. 

As one can see, the normalization integral coincides with the
normalization of the pion electromagnetic form factor,
\begin{eqnarray}
\int_0^1 dx q(x) = \int_0^1 dx \bar q (x) = F_\pi^{\rm em} (0) = 1 .
\end{eqnarray}   
In addition, we have the
crossing property, 
\begin{eqnarray} 
\bar q (x) = q (1-x).
\end{eqnarray}

The $k_\perp$-unintegrated parton distribution can be shown to be
equal to 
\begin{eqnarray}
q(x,k_\perp)=\frac{N_c}{4 \pi^3 f_\pi^2} \int d\omega \rho(\omega) 
\frac{\omega^2}{k_\perp^2+\omega^2} \theta(x)\theta(1-x) , \nonumber
\\ \label{psf7}
\end{eqnarray}
which is the same form as in Eq.~(\ref{pionlcwf}),
hence at the working scale of the model, $Q_0$, 
one has the interesting relation
\begin{eqnarray}
q(x,k_\perp)= \bar q(1-x, k_\perp ) = \Psi(x,k_\perp). \label{qpsf}
\end{eqnarray}
valid in our model in the chiral limit. A similar identity has also
been found in the Nambu--Jona-Lasinio model~\cite{Ru01}. Combining Eq.~(\ref{qpsf})
with the anomaly condition (\ref{consist}), we get the following
normalization for the unintegrated parton distribution at $k_\perp=0$, 
\begin{eqnarray}
q(x,0_\perp) =\frac{N_c}{4 \pi^3 f_\pi^2}  . 
\end{eqnarray} 
Finally, via integrating with respect to $k_\perp$ the following identity
between the PDF and the PDA is obtained at the scale $Q_0$: 
\begin{eqnarray}
q(x)= \varphi_\pi(x)  .
\label{eq:pdf=pda}
\end{eqnarray} 
This relation holds also in the Nambu--Jona-Lasinio model with the Pauli-Villars
regularization~\cite{RB02}. 

\subsection{Momentum sum rule}

In our formalism the pion expectation value of the energy-momentum tensor 
(\ref{eq:en-mom} ) is
\begin{eqnarray}
&& \langle \pi^a (q) | \theta^{\mu \nu} (0) | \pi^b (q) \rangle  = - N_c \int
d\omega \rho(\omega) \left ( {\omega\over f_\pi} \right)^2 \times  \nonumber \\
&& \int {d^4 k \over
(2\pi)^2 } {\rm Tr} \left[ \gamma_5 \tau_a
{1\over \slashchar{k}+\slashchar{q}-\omega} \gamma_5 \tau_b {1\over
\slashchar{k}-\omega} \times \right.  \\ && \left. \left\{ \frac12 (k^\mu \gamma^\nu + k^\nu
\gamma^\mu) - g^{\mu \nu} ( \slashchar{k}-\omega) \right\} {1\over
\slashchar{k}-\omega} \right]. \nonumber
\end{eqnarray} 
Though the use of the spectral conditions we get, for on-shell
massless pions, 
\begin{eqnarray}
\langle \pi^a(q) | \theta^{\mu \nu} | \pi^b(q) \rangle = 2 \left[ q^\mu q^\nu +
g^{\mu \nu} \frac2{N_f} \frac{B}{f_\pi^2} \right] \delta^{ab}.
\end{eqnarray} 
The connected piece becomes
\begin{eqnarray}
\langle \pi^a(q) | \theta^{\mu \nu} | \pi^b(q) \rangle_C  = 2  q^\mu q^\nu .
\end{eqnarray} 
Thus, the quarks carry all momentum of the pion, as it should be,
since there are no other degrees of freedom in the model. As it is
well known~\cite{Ja85,Ellis:1976eh}, the matrix element of the energy
momentum tensor coincides with the first moment of the PDF
\begin{eqnarray}
\int_0^1 dx \,x q(x)  = \int_0^1 dx \, x \bar q(x) = \frac12  .
\end{eqnarray} 
Actually, this property is a simple consequence of the crossing
property $\bar q (x) = q(1-x) $ and the normalization condition.

\subsection{QCD evolution} 

The QCD evolution of the constant pion structure function has been
treated in detail in previous works \cite{DR95,DR02} at LO and NLO
order. Nevertheless, in order to make the paper more self-contained we
present here the main points arising from that discussion. The
previous subsections yield the form of the leading-twist contribution
to the pion structure function at a given renormalization point, $Q_0$. 
The QCD radiative corrections generate logarithmic scaling violations,
which can be included in our model by the DGLAP equations~\cite{AP77}. In
particular, the non-singlet contribution to the energy momentum tensor
evolves as
\begin{eqnarray}
\frac{ \int dx \,x q(x, Q) } { \int dx \, x q(x,Q_0) } = \left(
\frac{\alpha(Q)} {\alpha(Q_0) } \right)^{\gamma_1^{(0)} / (2
\beta_0) } \quad , \qquad
\end{eqnarray} 
where $ \gamma_1^{(0)}$ and $\beta_0$ are given in Eq. (\ref{gambe}).  In
Ref.~\cite{SMRS92} it was found that at $Q^2 = 4 {\rm GeV}^2 $ the valence
quarks carry $47 \pm 0.02$\% of the total momentum fraction in the
pion. Downward LO evolution yields that at the scale 
\begin{eqnarray}
Q_0 = 313_{-10}^{+20} {\rm MeV}  
\label{eq:mu0_dis} 
\end{eqnarray} 
the quarks carry $100\%$ of the momentum. The agreement of the evolved
PDF~\cite{DR95,DR02} with the data analysis~\cite{SMRS92} is quite
impressive. Equation~(\ref{eq:pdf=pda}) has been shown in Ref.~\cite{RB02}
to produce a very interesting integral equation relating the evolved PDF and
PDA, valid at the leading order QCD evolution.

\section{Gasser-Leutwyler coefficients}

The gauge technique provides a way to deal with open quark lines, an
advantage over traditional chiral quark models, but it is
unnecessarily complicated when dealing with processes with closed
quark lines. For such a situation the effective action approach
provides a much more efficient calculational tool. It also yields a
closer connection to previous approaches such as bosonized versions of
the Nambu--Jona-Lasinio model.  The one-quark-loop effective action that incorporates
the quark-pion coupling obeying the Goldberger-Treiman relation
(\ref{eq:gt}) can be written in the form
\begin{eqnarray}
S&=&-i N_c \int d^4x \int d\omega \rho(\omega)\times \nonumber \\
&&{\rm Tr}\log \left [ i \slashchar{\partial} - \omega 
\exp \left ( i \gamma_5 \tau_a \phi_a(x)/f_\pi \right ) \right ] . 
\label{action}
\end{eqnarray} 
This form is manifestly chirally symmetric, with $\phi$ denoting the
non-linearly realized pion field. Note the formal similarity with a
generalized Pauli-Villars regulator. One may evaluate the
Gasser-Leutwyler coefficients~\cite{GL84,GL85} through the use of
standard derivative expansion techniques~\cite{Chan:1986jq} in a similar
fashion as done in Ref.~\cite{Ru91,SRG92}. With the (\ref{rho0})
condition imposed, the calculation is equivalent to standard
quark-model calculations with the cut-off removed. The resulting
values of the Gasser-Leutwyler coefficients are
\begin{eqnarray}
\bar l_1&=& -N_c , \nonumber \\ 
\bar l_2&=& N_c .\label{GL}
\end{eqnarray}
Other low energy constants, such as $\bar l_3$ and $\bar l_4 $ require
a specification of explicit chiral symmetry breaking within the quark
model. External gauge fields may also be coupled resulting in
predictions for $\bar l_5 $ and $\bar l_6$, although any choice
reflects a particular selection of transverse pieces. This and related
issues are postponed for future studies. It is nevertheless
interesting to anticipate here a dimensional argument which shows why
the relevant spectral condition for the terms involving fourth-order derivatives is,
precisely, $\rho_0 = \int d \omega \rho(\omega) = 1 $. In the case of the
$\pi\pi $ scattering in the chiral limit, we have a box diagram with
four quark propagators, $ i/(\p - \omega)$, and four external pion
lines, each contributing a factor of $\omega / f_\pi $, due to the
Goldberger-Treiman relation (\ref{GTrel}).  If we are after the coefficient with
four derivatives we need four additional powers of momenta in the
denominator, which we may account for by squaring the fermion
propagator.  Thus, in obvious dimensional notation we have after
adjusting the dimensions,
\begin{eqnarray}
\frac{\bar l}{f_\pi^4}(\partial \phi )^4 &\sim& \int \rho(\omega)
d\omega \int \frac{d^4 p}{(2\pi)^4} \left( \frac{\omega}f_\pi
\right)^4 \left( \frac{i}{p^2 -\omega^2} \right)^4 (\partial \phi) ^4
\nonumber \\ &\sim & \int \rho(\omega) d\omega
\frac1{f_\pi^4}(\partial \phi )^4
\end{eqnarray}    
due to the fact that the dimensions of the convergent integral are set
by the spectral mass $\omega$. This shows that the terms of dimension
four in the effective Lagrangian are proportional to $\rho_0$.

\section{Vector-meson dominance} \label{sec:VMD} 

Up to now, our considerations have been made for a {\it general}
spectral function fulfilling a set of properties regarding their
moments and log-moments. It is quite natural to ask whether such a
function exists and what are the phenomenological consequences of
making specific ans\"atze for this function.  In this section we
construct explicitly the spectral function using the phenomenological
guidance of the previous sections. 

Some interesting consequences and insight may be obtained in the
present chiral quark model if the vector-meson dominance of the 
pion form factor is
assumed,
\begin{eqnarray} 
F_V (t) &=& \frac{M_V^2}{M_V^2+t}. \label{VMD}
\end{eqnarray} 
with $M_V$ denoting the $\rho$-meson mass. This form fits the recent
data \cite{Volmer:2000ek} remarkably well. As will be shown below, the
model for the spectral function becomes explicit and further
interesting results may be obtained.

\subsection{Vector-meson dominance in the spectral approach} 

The vector form factor obtained in Eq.~(\ref{eq:ffpi}) reads, through the use
of the Feynman parameterization~(\ref{eq:ifeynman}),  
\begin{eqnarray}
&& F_V (t) \equiv F_\pi^{\rm em}(t)= \\
&& -\frac{N_c}{4\pi^2 f_\pi^2} \int d\omega \rho(\omega) \omega^2 \int_0^1 dx
\log\left[\w^2+x(1-x) t \right] . \nonumber 
\end{eqnarray} 
If we make a series expansion in $t$, the integral in $x$ can be
carried out order by order, hence 
\begin{eqnarray}
F_V (t) &=& 1 + \frac{N_c}{4\pi^2 f_\pi^2 } \sum_{n=1}^\infty \int d \w
\rho(\omega) \omega^2 \nonumber \\ &\times&  \int_0^1 dx [x(1-x)]^n
\frac{(-1)^n}{n}\left(\frac{t}{\omega^2}\right)^n \label{FVexp} \\ &=&
\frac{N_c}{4\pi^2 f_\pi^2 } \sum_{n=1}^\infty \rho_{2-2n} \frac{ 2^{-2n-1}
\sqrt{\pi} \Gamma(n+1)}{n \Gamma(n+3/2)} \left(-t\right)^n . \nonumber
\end{eqnarray} 
As one can see, in our model the pion form factor
is encoded in the negative even moments. 
Through vector meson dominance we get immediately, by comparing 
Eq.~(\ref{FVexp}) to the expansion of Eq.~(\ref{VMD}), the following
identification
\begin{eqnarray} 
\rho_{2-2k} &=& \frac{2^{2k+3} \pi^{3/2} f_\pi^2 }{N_c M_V^{2k} }
\frac{k \, \Gamma(k+3/2)}{ \Gamma(k+1)},   \nonumber \\
&& k=1,2,3,\dots  \label{evmom}
\end{eqnarray} 
In particular, the normalization condition, $\rho_0=1$, yields
\begin{eqnarray}
M_V^2 = \frac{24 \pi^2 f_\pi^2 }{N_c} .
\label{Mvfpi}
\end{eqnarray} 
This relation is usually obtained when matching chiral quark models to
the vector-meson dominance and appears all over the literature,
yielding a quite reasonable estimate for the $\rho$ meson mass, $M_V=
826~{\rm MeV}$ with $f_\pi=93$~MeV, and $M_V= 764~{\rm MeV}$ with
$f_\pi=86$~MeV in the chiral limit.  

The interesting and remarkable
point about Eq.~(\ref{evmom}) is that even though we have determined
the negative even moments of the spectral function, the positive even
moments, obtained by analytic continuation in the index $n$,
unexpectedly but most desirably, 
fulfill the spectral conditions (\ref{rhon}) for the
positive moments due to the fact that $\Gamma(n)$ has single poles at
non-positive integers, $n=0,-1,-2, \dots$ Hence
\begin{eqnarray}
\rho_{2n} = 0, \qquad n=1,2,3 \dots  
\end{eqnarray} 
Thus, it makes sense to evaluate the log-moments (\ref{rholog}), since the
absence of
 dimensional transmutation is guaranteed. The log-moments
are most easily evaluated
 by analytically continuing the moments 
to the complex $n$-plane. We then have
\begin{eqnarray}
\rho'_{n}&=&\int d\omega \omega^n \log(\omega^2) \rho(\omega) = 
\left . 2 \frac{d}{dz} \int d\omega \omega^z \rho(\omega) \right |_{z=n} \nonumber \\
&=&\left . 2 \frac{d}{dz} \rho_z \right |_{z=n},
\label{drhodz}
\end{eqnarray}
and
\begin{eqnarray}
\rho'_{2n} &=& \left(-\frac{M_V^2}{4} \right)^n 
\frac{ 
\Gamma (n ) \,  \Gamma \left( \frac52-n \right)}{\Gamma(\frac52)}, 
\nonumber \\ && \qquad \qquad \qquad n=1,2,3 \dots ,
\label{eq:log-vmd}
\end{eqnarray}   
where we have used Eq. (\ref{Mvfpi}).
The first few values are 
\begin{eqnarray}
\rho'_{2} &=& - \frac{4 f^2 \pi^2}{N_c}, \nonumber  \\ 
\rho'_{4} &=&  \frac{2 f^2 M_V^2 \pi^2}{N_c}, \nonumber  \\
\rho'_{6} &=&  \frac{2 f^2 M_V^4 \pi^2}{N_c}.   
\end{eqnarray} 
Since $\rho'_{2}$ and $\rho'_{4}$  determine $f_\pi$ and $B$,
respectively, see Eq.~(\ref{eq:w2log},\ref{eq:w4log}), we may write 
the following interesting relation coming from the vector-dominance model
and the spectral approach: 
\begin{eqnarray} 
B&=& -\frac{9 \pi^2 f_\pi^4}{N_c} = -\frac{N_c M_V^4}{64 \pi^2} \nonumber \\
&=&- (202-217~{\rm MeV} )^4 .
\end{eqnarray} 
The uncertainty stems only from using either $f_\pi$ or $M_V$ as
input. Our value agrees within errors with the estimate (\ref{Bqcd}).

\subsection{Inverse problem} 

Although for practical calculations the moments seem to contain the
relevant information that can be used in  practical applications, it
is nevertheless very interesting to write down an
 explicit formula
for the spectral function. The mathematical problem is then to
invert the formula
\begin{eqnarray}
\rho_{2n} &=& \int_C d\omega \omega^{2n} \rho_V (\omega), \label{2nmom}
\end{eqnarray}  
where the moments are given by Eq.~(\ref{evmom}). The solution to
the problem is given by the following surprisingly simple function,
\begin{eqnarray}
\rho_V (\omega) &=& \frac{1}{2\pi i} \frac{1}{\omega}
\frac{1}{(1-4\omega^2/M_V^2)^{d_V}},  \nonumber \\
\label{rhoVdv}
\end{eqnarray} 
with 
\begin{equation}
d_V=5/2,
\end{equation} 
at which case we have  
\begin{eqnarray}
\!\!\!\!\!\!\!\!\!\! \rho_V (\omega) = \frac{1}{2\pi i} \frac{3 \pi^2 M_V^3 f_\pi^2 }{4 N_c}
\frac{1}{\omega} \frac1{(M_V^2/4-\omega^2)^{5/2}}. 
\label{v52}
\end{eqnarray} 
The function $\rho_V(\omega)$ has a single pole at the origin and branch cuts starting
at $\pm$ half the meson mass, $\omega=\pm M_V/2$. The contour for
computing the spectral moments is depicted in
Fig.~\ref{fig:contour}. The contributions encircling the branch points
cancel provided $ d_V $ is half-integer. The
contribution at infinity cancels, and only the residues at the origin
contribute. One may explicitly verify with no difficulty that the use
of Eq.~(\ref{rhoVdv}) in Eq.~(\ref{2nmom}) with the contour of
Fig. \ref{fig:contour} reproduces Eq.~(\ref{evmom}).

With help of the explicit formula (\ref{rhoVdv}) several
interesting features may be pointed out. The spectral function is
genuinely complex and it is defined on a complex contour $C$. This
precludes positivity conditions 
\footnote{If one extrapolates $\rho_V(\omega)$ 
to the real axis, as suggested by the standard Lehmann
representation, Eq.~(\ref{eq:leh_rep1}), there appear, for $ d_V \ge 1$, 
end-point non-integrable singularities at the branch points $\omega=\pm M_V /2$. 
If one insists on a {\em real} spectral function, one
may do so but then one has to proceed by analytic continuation in
$d_V$, or derivation with respect to $M_V^2$ after computing the
$\omega$-integral. Alternatively, this is equivalent to a
distributional interpretation of the spectral function $\rho (\omega)$
and its (generalized) derivatives using the well known distribution $
x_+^\alpha $ (see e.g. the classical work \cite{GS64} for a rigorous
discussion).}.

It is also interesting to note that in the limit $M_V \to \infty $ one
gets 
\begin{eqnarray}
\rho_V (\omega) & \to & \frac{1}{2\pi i} 
\frac{1}{\omega} , 
\end{eqnarray} 
{\em i.e.} the massless free theory. This suggests a multiplicative effect
of chiral symmetry breaking on the spectral function. 
Finally, Eq.~(\ref{rhoVdv}) cannot be interpreted as a constituent
model for which one essentially has a pole at $\omega = M$ in the complex
plane (or, equivalently, a $ \delta(\omega -M)$), with $M$ 
denoting the constituent quark mass. 

\begin{figure}[]
\includegraphics[width=8.5cm]{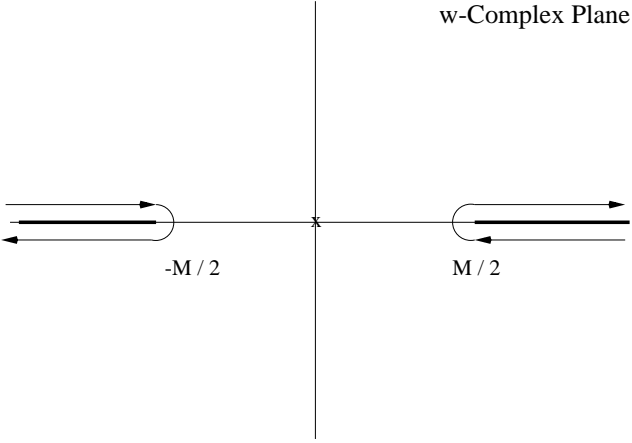} 
\caption{The contour in the complex $\omega$-plane for the spectral functions
in the meson dominance model. The quantity $M$ denotes the generic meson mass, $M_V$
for the vector channel, and $M_S$ for the scalar channel. 
The branch cuts start at $\omega = \pm M/2$. The pole at $\omega=0$ occurs
for the vector case only.}
\label{fig:contour}
\end{figure}

\section{Quark propagator in the meson dominance model} \label{sec:q_vmd}

\subsection{Scalar spectral function}

In the construction of the vector spectral function we have used the
vector-meson dominance principle, which has a firm phenomenological
justification. For the case of the scalar spectral function we will
proceed differently, more heuristically. First, we will propose its form in an analogy to
the form of $\rho_V$. Then, in Sect. \ref{sec:lattice} we will confront
our hypothesis to the recent lattice data on the quark
propagator~\cite{Bowman:2002bm,Bowman:2002kn}.

The scalar function has to satisfy the conditions (\ref{rhon}) at
odd positive values of $n$.  The analysis of the previous section
suggests the following form 
\begin{eqnarray}
\rho_S (\omega)= \frac{1}{2\pi i} \frac{16(d_S-1)(d_S-2) \rho'_3}
{M_S^4 (1-4\omega^2/M_S^2)^{d_S}}, \nonumber \\ \label{rhoSds}
\end{eqnarray} 
where the normalization is chosen in such a way that the third log
moment, $\rho'_3 =- 4\pi^2 \langle \bar q q\rangle/N_c $, is
reproduced. In other words, we fix the normalization with the quark
condensate. The admissible values of $d_S$ are half-integer, since
only then the integration around the half-circles at the branch
points in Fig. \ref{fig:contour} vanishes.  In Sect.~\ref{sec:lattice}  
%\ref{Bowman:2002bm,Bowman:2002kn,}
the preferred value will turn out to be 
\begin{equation}
d_S=5/2.
\end{equation}

One may verify
that the integration with the prescription of Fig. \ref{fig:contour}
yields
\begin{eqnarray}
&& \rho_{2k-1} = \int_C d\omega \omega^{2k-1} \rho_S (\omega) \label{2km}
\\ &&= - \frac{4^{3-k} \pi^2 \langle \bar q q
\rangle}{M_S^{-2k+4} N_c} \frac{\Gamma(d_s-k)}{\Gamma(d_S-2)
\Gamma(1-k)}, \nonumber  
\end{eqnarray}  
and conditions (\ref{rhon}) become satisfied for odd positive $n=2k-1$.
The analytic structure of $\rho_S(\omega)$ is similar to the case of  $\rho_V(\omega)$,
except for the absence of the pole at $\omega=0$. 

For completeness, we list the result for even and odd negative
moments,
\begin{eqnarray}
\rho_{-2n} &=& \left ( \frac{M_V^2}{4} \right )^{-n} \frac{\Gamma(d_V+n)}{\Gamma(d_V)
\Gamma(n+1)}, \\ 
\rho_{-2n-1} &=& - \left( \frac{M_S^2}{4} \right )^{-n-2} 
\frac{4 \pi^2 \langle \bar q q \rangle}{N_c} \frac{\Gamma(d_s+n)}{\Gamma(d_S-2)
\Gamma(n+1)}, \nonumber \\
&& n=0,1,2, \dots \nonumber
\end{eqnarray} 
and for the positive even and odd log-moments, 
\begin{eqnarray}
\rho'_{2n} &=& \left(-\frac{M_V^2}{4} \right)^n \frac{ \Gamma(d_V-n)
\Gamma(n)} {\Gamma(d_V)}, \\ 
\rho'_{2n-1} &=&
-\left(-\frac{M_S^2}4\right)^{n-2} \frac{4 \pi^2 \langle \bar q q
\rangle}{N_c}\frac{ \Gamma(d_S-n) \Gamma(n)} {\Gamma(d_S-2)}, \nonumber \\
&& n=1,2,3,\dots \nonumber
\end{eqnarray} 
The value $d_V=5/2$ should be used for the vector dominance model.

\subsection{Quark propagator}  

A straightforward calculation with Eq.~(\ref{rhoVdv},\ref{rhoSds})
yields the $A(p^2) $ and $B(p^2)$ functions of Eq.~(\ref{ABdef}), namely 
\begin{eqnarray}
A(p^2) &=& \frac1{p^2} \left[ 1 - \frac1{(1-4p^2/M_V^2)^{d_V}} \right],
\nonumber \\ 
B(p^2) &=& \frac{64 (d_S -2) (d_S-1) \pi^2 \langle \bar q q\rangle }
{M_S^4 N_c (1-4p^2/M_S^2)^{d_S}} \label{AB}.
\end{eqnarray} 
We note that the apparent pole in $A(p^2)$ is canceled when the expression
in brackets is expanded, and both
functions {\em have no poles in the whole complex plane}. The functions
(\ref{AB}) have branch cuts starting at $p^2=4 M^2$, where $M$ is the
relevant mass. 

The absence of poles, achieved in a rather natural
fashion in our approach, is very appealing, but not completely
surprising {\it a posteriori}. In local chiral quark models, where the
propagator is usually assumed to be a meromorphic function (with a
pole at the constituent quark mass), meson vertex functions naturally
inherit the discontinuity structure implied by the Cutkosky rules and
unitarity. In our case, it is the meson form factor which is
taken to be a meromorphic function through the VMD model; unavoidably,
the quark propagator must have no poles and 
a certain cut structure conspiring with the
unitarity at the one-loop level in order to produce such a form factor with no
cuts.

Alternatively, instead of $A$ and $B$ one may consider the more customary mass function,
$M(p^2)$, and the wave function renormalization, $Z(p^2)$, given
by Eq.~(\ref{Snl}). They can be written as
\begin{eqnarray} 
\frac{M(p^2)}{M_0} &=& \frac{ 4 d_V p^2}{M_V^2}\frac{
\left(\frac{M_S^2}{M_S^2 -4 p^2} \right)^{d_S}}{
\left(\frac{M_V^2}{M_V^2 -4 p^2} \right)^{d_V} -1} \label{MZmod} 
\\ Z(p^2) &=& 1 -
\left(\frac{M_V^2}{M_V^2 -4 p^2} \right)^{d_V} \\ &+&
\frac{16 d_V^2 M(0)^2 p^2 }{M_V^4} \frac{ \left(\frac{M_S^2}{M_S^2 -4
p^2} \right)^{2d_S}}{ \left(\frac{M_V^2}{M_V^2 -4 p^2} \right)^{d_V}
-1}, \nonumber
\end{eqnarray} 
where $M_0=M(0)$ is the value of the mass at the origin. We find
\begin{eqnarray} 
M_0 &=& -\frac{16 (d_S-1)(d_S-2) M_V^2 \pi^2 \langle \bar q q \rangle
}{d_V M_S^4 N_c}, \nonumber \\ Z(0) &=& \frac{ 4 d_V M(0)^2}{M_V^2} .
\end{eqnarray} 
At high Euclidean momenta, $Q^2=-p^2 \to \infty $,  we obtain
\begin{eqnarray}
&& \!\!\! M(Q^2) = \frac{d_V M_0 M_S^2}{M_V^2}\left(\frac{M_S^2}{ 4 Q^2}
\right)^{d_S-1} + \dots , \nonumber \\ 
&& \!\!\! Z(Q^2) = 1 -
\left(\frac{M_V^2}{4 Q^2} \right)^{d_V} - \frac{ 4 d_V^2 M(0)^2
M_S^2 }{M_V^4} \left(\frac{M_S^2}{4 Q^2} \right)^{2d_S-1}  \nonumber \\
&& +\dots .
\end{eqnarray} 
We note that for half-integer $d_S$ the tail of $M$ contains odd
powers of $Q$, 
\begin{eqnarray}
M(Q^2) \sim \frac{1}{(Q^2)^{d_S-1}},
\end{eqnarray}
and for $d_S=5/2$ 
 drops as $1/Q^3$, for $d_S=7/2$ as
$1/Q^5$, {\em etc.} The wave-function normalization, $Z(Q)$, has the
correct  asymptotic behavior, $Z(Q) \sim 1$.  

Given the cut structure of the functions $A(p^2)$ and $B(p^2)$ we may
look back at the high energy expansion (\ref{ShighQ}). According to
the spectral conditions (\ref{rhon}) one would deduce from the high
energy behavior that the full propagator coincides with the free
one. The puzzle is resolved by realizing that the branch cut running
from $\pm M /2 $ to $\pm \infty $ implies a fractional power behavior,
and hence the function cannot be represented by a power series
expansion around infinity. From this point of view Eq.~(\ref{ShighQ})
just expresses the fact that the integer power coefficients are
exactly zero, as clearly follows from Eq.~(\ref{AB}).

\subsection{Comparison to the lattice data} \label{sec:lattice} 

\begin{figure}
\includegraphics[width=8.8cm]{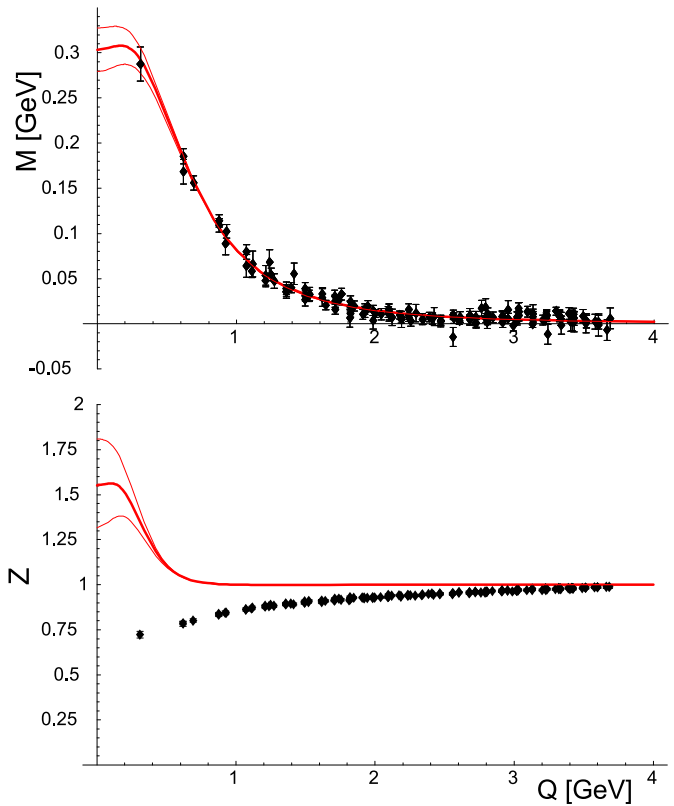} 
\caption{The dependence of the quark mass $M$ (top) and the wave
function normalization $Z$ (bottom)  on the Euclidean momentum $Q$
obtained from the meson dominance model with $d_S=5/2$.  The lattice data,
extrapolated to the chiral limit for the case of $M$,  and taken at
the current quark mass $\simeq 59$~MeV for the case of $Z$,  are
taken from Ref.~{\protect{\cite{Bowman:2002bm,Bowman:2002kn}}}.  The
thicker lines correspond to optimum parameters of Eq.~(\ref{optim}),
while the thin lines  indicate the uncertainty at the
one-standard-deviation level.}
\label{fig:MZ}
\end{figure}

The decrease of the quark mass as $1/Q^3$ at large Euclidean momenta is
favored by recent lattice
calculations~\cite{Bowman:2002bm,Bowman:2002kn}, where the fit to the
functional form $c/((Q^2)^b+(\Lambda^2)^b)$ is best when the parameter
$b$ is close to 3 \footnote{Obviously, $M(p^2)$ and $Z(p^2)$ are gauge-dependent
quantities. The data of Ref.~\cite{Bowman:2002bm,Bowman:2002kn} 
correspond to the Landau and Laplacian gauges. A natural question arises where 
our results incorporate the choice of the QCD gauge at the microscopic level. 
We notice two sources of ambiguities in our approach: 
the arbitrariness of the transverse terms in the solutions to the 
Ward-Takahashi identities, as well as the choice of the scalar quark spectral function. 
It remains to be seen how these choices reflect the underlying QCD gauge.}. 
We have performed a $\chi^2$ fit for $M(Q)$ of
Eq.~(\ref{MZmod}) to the data of
Ref.~\cite{Bowman:2002bm,Bowman:2002kn}. These data, for the case of
$M$, have been extrapolated to the chiral limit. We  have treated
$M_S$ and  $M_0$ as free parameters. The $\chi^2$ fit
results in the following optimum  values,
\begin{eqnarray}
M_0&=& 303 \pm 24~{\rm MeV}  , \nonumber \\
M_S&=& 970 \pm 21~{\rm MeV}, \label{optim}
\end{eqnarray}
with the optimum value of $\chi^2$ per degree of freedom equal to 0.72. 
The corresponding value of the quark condensate is    
\begin{eqnarray}
\langle \bar q q \rangle &=& -(243.0^{+0.1}_{-0.8}~{\rm MeV})^3. \label{opti}
\end{eqnarray}
The functions $M(Q)$ and $Z(Q)$, evaluated at optimum parameters, are
shown in Fig. \ref{fig:MZ} with thick lines. The thin lines 
indicate the uncertainty at the one-standard-deviation level. The
agreement  with the data is very good for the case of $M$, and the
$1/Q^3$ fall-off is clearly seen. In fact, fitting of the model with
values of $d_S$ higher than $5/2$, which results in faster
asymptotic decrease, results in a much worse agreement with the
data. As seen from the bottom part of Fig. \ref{fig:MZ},  for the case
of $Z$ the agreement is not very good, but we should keep in mind the
simplicity of the present model and the freedom in the scalar channel.  For
instance, the scalar spectral function can be multiplied,
without loosing any of the  general requirements, by an entire
function.  We also wish to stress that the optimum value of $\langle
\bar q q \rangle$ obtained by  fitting the model formulas to the
lattice data agrees with the estimate (\ref{opti}). 

\begin{figure}
\includegraphics[width=8.5cm]{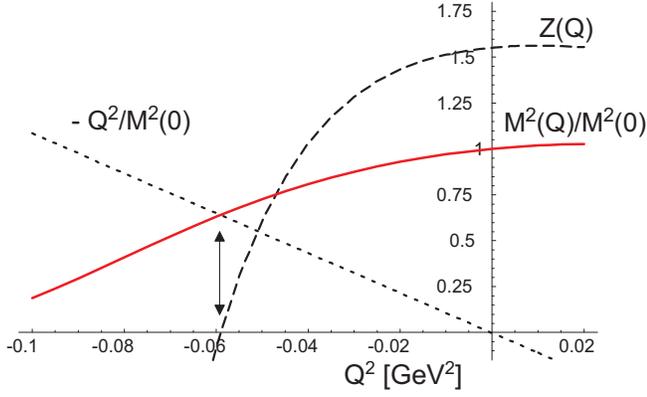} 
\caption{Square of the quark mass $M^2$ (solid line), minus the square
of the Euclidean momentum $Q^2$ (dotted line),  and the wave
function normalization $Z$ (dashed line), plotted as functions of the
square of Euclidean momentum, $Q^2$.  The arrow indicates the point
where $M^2=-Q^2$, but where also $Z=0$. As the result, the quark 
propagator has no pole. From the analyticity properties of $A$ and $B$
it follows that the quark propagator has no poles in  the whole
complex-$Q^2$ plane.}
\label{fig:MZlowQ}
\end{figure}

It is interesting to note that even though the mass function,
$M(p^2)$, presents a pole for time-like momenta, {\em i.e.} there
exists a solution to the equation $M(p^2)-p^2=0$, it does not
correspond
 to a physical particle. This is because the
normalization $Z(p^2)$ also vanishes for the same value
 of
$p^2$. This in fact is just a manifestation of the analyticity 
properties of $A(p)$ and $B(p)$ discussed above. 
 Figure
\ref{fig:MZlowQ} shows the behavior of $M$ and $Z$ at low
momenta. Arrows indicate the positions of the
 alleged pole in $M$,
canceled by the zero of $Z$.

\section{Other predictions} \label{other} 

The explicit model for the quark spectral function $\rho(\omega)$
allows for very simple and efficient evaluation of further interesting
quantities. There is a whole bunch of predictions, from which we only list
a few. The results decouple into those involving the vector spectral
function, $\rho_V(\omega)$, and the 
 scalar spectral function,
$\rho_S(\omega)$. As stresses throughout the paper, all predictions are made for the model working
scale $Q_0$, and the QCD evolution is needed if comparison to high-energy data
is desired.

\subsection{Pion transition form factor} 

According to the vector meson dominance model of the pion form factor,
the transition form factor (\ref{eq:tffpi}) becomes
\begin{eqnarray}
&& F_{\pi \gamma^\ast \gamma^\ast}(Q^2 , A) = \frac{2 f_\pi}{ A N_c} \frac1{Q^2} \log \left[\frac{2M_V^2 +
 (1+A) Q^2 }{2 M_V^2 + (1-A) Q^2  }\right]  \nonumber \\ &+&
\frac{16 f_\pi M_V^2 }{N_c [4 M_V^4 + 4Q^2 M_V^2 + (1-A^2) Q^4 ]} 
\end{eqnarray} 
As discussed in Sect. \ref{anomff}, this expression provides the twist
expansion, with the result of Eq.~(\ref{phin}), but with VMD the log-moments
 of Eq.~(\ref{eq:log-vmd}). The
analysis and comparison to the data of Ref.~\cite{CLEO98} requires the QCD
evolution of the higher twist components. This point will be analyzed
elsewhere.

\subsection{Pion light-cone wave function \label{sec:lc}}

Next, we use Eq.~(\ref{v52}) in the expression for the pion 
light-cone wave function (\ref{pionlcwf}), with the result
\begin{eqnarray}
\!\!\!\! \Psi(x,k_\perp)=\frac{3M_V^3}{16\pi(k_\perp^2+M_V^2/4)^{5/2}}
 \theta(x)\theta(1-x). \label{Psimod}
\end{eqnarray}
Passing to the impact-parameter space with the Fourier-Bessel
transform yields
\begin{eqnarray}
\Psi(x,b)& \equiv & 2 \pi \int_0^\infty k_\perp d k_\perp \Psi(x,k_\perp) J_0(k_\perp b) \nonumber \\
&=& \left (1+\frac{b M_V}{2} \right )
 \exp\left ( - \frac{M_V b}{2} \right ) \times \nonumber \\ &&
 \theta(x)\theta(1-x).  \label{Psibmod}
\end{eqnarray}
The expansion at small $b$ yields
\begin{eqnarray}
\Psi(x,b)&=& \left( 1-\frac{M_V^2 b^2}{8}+\frac{M_V^3 b^3}{24}+\dots
\right)\nonumber \\ &\times& \theta(x)\theta(1-x).
\label{Psibexp}
\end{eqnarray}
The average transverse momentum squared is equal to  
\begin{eqnarray}
\langle k_\perp^2 \rangle &\equiv& \frac{\int d^2 k_\perp \, k_\perp^2 \Psi(x,k_\perp)}
{\int d^2 k_\perp \Psi(x,k_\perp)}=-4\left.\frac{d\Psi(x,b)}{db^2}\right|_{b=0} \nonumber \\
&=& \frac{M_V^2}{2}, \label{kperp2}
\end{eqnarray}
which numerically gives $\langle k_\perp^2 \rangle=(544~{\rm MeV})^2$ (all at the model working scale
$Q_0$). This value is not far from the result of the Nambu--Jona-Lasinio model with the Pauli-Villars regularization,
which produces $\langle k_\perp^2 \rangle=(613~{\rm MeV})^2$ \cite{RB02}. 
The estimates from QCD sum rules yield smaller values:  
one gets $(316 {\rm MeV})^2$ based on Ref.~\cite{Zh94}, and 
$(333 \pm 40 {\rm MeV})^2$ based on Ref.~\cite{BI82}. One should note that these estimates are at 
$Q_0$ of the order of 1~GeV, hence the QCD evolution is needed to compare to our model results \cite{RB02}.

\subsection{Unintegrated pion structure function \label{sec:unps}}

Due to relation (\ref{eq:pdf=pda}), exactly the same formulas as in the
previous section hold for the  unintegrated quark distribution
function in the pion. Hence, we find 
\begin{eqnarray}
q(x,k_\perp)&=&\overline{q}(1-x,k_\perp) \label{qqmod} \\
&=&\frac{3M_V^3}{16 \pi(k_\perp^2+M_V^2/4)^{5/2}}
 \theta(x)\theta(1-x), \nonumber
\end{eqnarray}
and the analogs of Eqs. (\ref{Psibmod}-\ref{kperp2}).

\subsection{Non-local quark condensate}

The applications of Sect. \ref{sec:lc} and \ref{sec:unps} involved the
vector spectral density. The scalar spectral density (\ref{rhoSds})
enters the evaluation of the nonlocal condensate of
Eq.~(\ref{qxw}). For $d_S=5/2$ we find immediately the interesting and
simple result
\begin{eqnarray}
Q(x)=\exp\left ( -\frac{M_S \sqrt{-x^2}}{2} \right ), \label{Qxmod}
\end{eqnarray}
where $x$ denotes the Minkowski coordinate.
For higher values of $d_S$ the expression is multiplied by a polynomial
in the $x^2$ variable. Note the nonanalyticity in the $x^2$ variable
in Eq.~(\ref{Qxmod}) as $x^2 \to 0$. In the present model the moments
of the quark condensate, Eq.~(\ref{eq:nlcon_mom}), are well defined
for $n < 2d_S - 4$, which means that for the preferred value of
$d_S=5/2$ we may only consider the quark condensate itself, but not
its moments. Nevertheless, the coordinate representation of the
non-local condensate makes sense and can be given for the whole range
of $x^2$.

\subsection{Quark propagator in the coordinate representation}

One may also pass to the coordinate representation for the $A$ and $B$
functions of Eq. (\ref{ABdef}), introducing
\begin{eqnarray}
-i \slashchar{x} A(x)&=&\int \frac{d^4 p}{(2\pi)^4} e^{-i p\cdot x} \p A(p), \nonumber \\
B(x)&=&\int \frac{d^4 p}{(2\pi)^4} e^{-i p\cdot x} B(p). 
 \label{ABx}
\end{eqnarray}
With straightforward algebra one finds
\begin{eqnarray}
A(x)&=&\frac{48+24M_V \sqrt{-x^2}-6M_V^2 x^2 + M_V^3
(-x^2)^{3/2}}{96\pi^2 x^4}\nonumber \\ &\times& e^{-M_V \sqrt{-x^2}/2}, \nonumber \\ B(x)&=&
\frac{2^{3-2d_S} \langle \overline{q} q \rangle}{N_c \Gamma(d_S-2)}
(M_S \sqrt{-x^2})^{d_S-2} K_{2-d_S}(M_S \sqrt{-x^2}/2), \nonumber \\
\label{ABexpl}
\end{eqnarray}
where $x$ denotes the Minkowski coordinate. 
We have used $d_V=5/2$. In the limit of low $x$ one 
recovers the result of the free theory, $A(x) \simeq 1/(2\pi^2 x^4)$. 
For the preferred value of $d_S=5/2$ we find that
\begin{eqnarray}
B(x)=\langle \overline{q} q \rangle /(4N_c) \exp(-M_S \sqrt{-x^2}/2),
\label{B52}
\end{eqnarray} 
in agreement with Eq. (\ref{Qxmod}).

\begin{table*}
\caption{Summary of conditions for the quark spectral function
$\rho(\omega)$.}
\begin{ruledtabular}
\begin{tabular}{ll}
Spectral condition & Physical significance \\ \hline
normalization & \\ \hline
$\rho_0=1$    & proper normalization of the quark propagator \\
              & preservation of anomalies \\
              & proper normalization of the pion distribution amplitude \\
              & proper normalization of the pion structure function \\
              & reproduction of the large-$N_c$ quark-model values of the Gasser-Leutwyler coefficients \\
              & relation $M_V^2=24\pi^2 f_\pi^2/N_c$ in the vector-meson dominance model \\ \hline
positive moments & \\ \hline
$\rho_1=0$    & vanishing quark mass at asymptotic Euclidean momenta, $M(Q^2)\to 0$ \\
$\rho_2=0$    & finiteness of the pion decay constant, $f_\pi$ \\
$\rho_3=0$    & finiteness of the quark condensate, $\langle \bar q q \rangle$ \\
$\rho_4=0$    & finiteness of the vacuum energy density, $B$ \\
$\rho_{n}=0, \qquad n=2,4,\dots $ & factorization in the twist expansion of vector amplitudes \\ 
$\rho_{n}=0, \qquad n=5,7,\dots $ & finiteness of  $\langle \bar q (\partial^2)^{(n-3)/2} q \rangle$ \\
 &  factorization in the twist expansion of the scalar pion form factor \\ 
\hline
negative moments & \\ \hline
$\rho_{-2}>0$    & positive value of the quark wave-function normalization at vanishing momentum, $Z(0)>0$ \\
$\rho_{-1}/\rho_{-2}>0$    & positive value of the  quark mass at vanishing momentum, $M(0)>0$ \\
$\rho_{-n}$ & low-momentum expansion of correlators \\ 
\hline
positive log-moments & \\ \hline
$\rho'_2 < 0$ & $f_\pi^2=-N_c/(4\pi^2)\rho'_2$ \\ 
$\rho'_3 > 0$ & negative value of the quark condensate, $\langle \bar q q \rangle = - N_c/(4\pi^2)\rho'_3$ \\
$\rho'_4 > 0$ & negative value of the vacuum energy density, $B = - N_c/(4\pi^2)\rho'_4$ \\
$\rho'_5 < 0$ & positive value of the squared vacuum virtuality of the quark, $\lambda^2_q=-\rho'_5/\rho'_3$ \\
$\rho'_n$     & high-momentum (twist) expansion of correlators \\
\end{tabular}
\end{ruledtabular}
\end{table*}

\section{Conclusion and final remarks}

In the present work we have developed a chiral quark model, which
tries to incorporate as many known features based on chiral symmetry
and the partonic quark substructure of hadrons as possible. This
approach, first unveiled in Ref.~\cite{Ru01}, should be
considered as a simple prototype of a construction which may be
certainly improved in many respects. Taking into account the fact that
the emerging picture is very encouraging, economic, and predictive, we
believe that applications and extensions of the model deserve a
thorough further investigation.

The key ingredient is the use of a generalized spectral representation
for the quark propagator combined with the gauge technique for
constructing the vertices involving quarks and currents. The
generalized Lehmann representation implies analyticity properties of
the quark propagator on the complex plane but positivity or reality
are abandoned. In particular, the analytic continuation from the
Euclidean to the Minkowski space back and forth becomes
straightforward.  We remind that this continuation is explicitly used
when computing hadronic matrix elements in momentum space. The
possibility of doing this continuation becomes extremely convenient when dealing
with calculations of soft matrix elements of high-energy processes. In
a purely Euclidean formulation one stays in the coordinate space and the
extraction of the parton distribution functions or amplitudes is often
limited in practice to the few lowest moments.

The conditions used to constrain the spectral function are collected
in Table I.  Finiteness and factorization enforce the vanishing of the
positive moments, while a number of available experimental observables
can be used to fix the values and signs of the negative moments and
the log-moments.

Instead of making a specific model for the propagator based on
extrapolations from the Euclidean region to the complex plane we
devise an infinite set of spectral conditions based on the requirement
that our model produces finite hadronic observables. As a result the
high energy behavior of certain matrix elements corresponds to a pure
twist expansion, with no logarithmic behavior. This is consistent with
the interpretation that the model is defined at a low renormalization
point $Q_0$. Thus, our model results should be considered as initial
condition for the QCD evolution, hence automatically incorporating the
correct high energy radiative corrections to hadronic observables. The
low renormalization point $Q_0$ is determined by the analysis of two
different processes: the pion transition form factor and the pion
structure functions. We find compatible values in the range $ Q_0 \sim
300 {\rm MeV} $, corresponding to $\alpha (Q_0) /(2 \pi) \sim 0.3
$. For the model scale, the leading twist contribution for the pion
structure function and the pion distribution amplitude coincide and
are equal to one, $q(x)=\varphi_\pi (x)=1$, regardless of the spectral
function. At the same time the correct anomalous form factor for the 
process $\pi^0 \to \gamma \gamma $ is obtained. This resolves the conflict between proper 
low-energy and high-energy normalization of the pion transition form
factor. The QCD evolution of both the PDF and PDA has been shown to
provide a very reasonable agreement with the data analysis.

The pion form factor does depend on the spectral function.  Further
interesting analytic relations can be obtained by determining the
even contribution to the spectral function from the requirement that
the pion form factor has the vector meson dominance form, which
is known to describe experimental data in the available momentum range 
very well. As a result, the
vector meson mass becomes proportional to the pion weak decay
constant. The vector contribution to the spectral function exhibits a
pole at the origin and non-integrable branch points at plus or minus
the vector meson mass. As a consequence the spectral function must be
interpreted as a distribution centered at the branch points.

The form of the spectral function suggested by the vector channel can
be also used with minor modifications for the scalar channel. This way
the full quark propagator can be studied and its analytic structure
analyzed: there are no poles all over the complex plane, only cuts
located at the branch points of the spectral function. Moreover, the
asymptotic behavior in the Euclidean region reflects this cut
structure by a half-integer power fall-off of the quark mass function
in the squared Euclidean momentum $Q^2$. The recent lattice data
\cite{Bowman:2002bm,Bowman:2002kn} set this half integer power to
$3/2$ quite unambiguously. Fixing this power leaves only the quark
mass function at the origin (the constituent quark mass) and the quark
condensate as free parameters. The fit to the data for the mass
function, $M(p^2)$, is very good and the values for both the
condensate and the constituent quark mass at $p^2=0$ agree with other
estimates. The fit to the quark wave-function normalization, $Z(p^2)$,
is not nearly as good as for $M(p^2)$, leaving room for
improvement.

Possible extensions of the general model involve the inclusion of more
general solutions of the Ward-Takahashi identities. The
meson-dominance model can be improved by studying more general forms
of the scalar quark spectral function. Moreover, the inclusion of
finite quark masses would allow the extension of the present model to
the complete pseudoscalar octet. These issues are under investigation.

\begin{acknowledgements}

We are grateful to the authors of
Ref.~\cite{Bowman:2002bm,Bowman:2002kn} for providing their
state-of-the art data for the quark propagator on the lattice.  We
thank Alexandr E. Dorokhov for a discussion on the pion electromagnetic form
factor.

This work is supported in part by funds provided by the Spanish DGI
with grant no. BFM2002-03218, and Junta de
Andaluc\'{\i}a grant no. FQM-225. Partial support from the Spanish
Ministerio de Asuntos Exteriores and the Polish State Committee for
Scientific Research, grant number 07/2001-2002 is also gratefully
acknowledged.
\end{acknowledgements}

\appendix 

\section{One-loop integrals}  \label{sec:appa}

\subsection{Two-point integral}

The two-point one loop integral regularized in $4+\epsilon$ dimensions is 
\begin{eqnarray}
&& \!\!\!\!\!\!\!\! I(q^2,\omega) = \frac1 i \int {d^4 k \over (2\pi)^4} {1\over k^2
- \omega^2 + i 0^+} {1\over (q-k)^2 - \omega^2 + i 0^+} \nonumber \\ 
&& \!\!\!\!\!\!\!\! =
\frac1{16\pi^2} \left ( 2+\sqrt{1-{4\omega^2\over q^2}} \log{ \sqrt{1-{4\omega^2\over
q^2}} -1 \over \sqrt{1-{4\omega^2\over q^2}} +1} \right )+ I(0,\omega), \nonumber \\
\end{eqnarray} 
with 
\begin{eqnarray}
I(0,\omega)=- \frac{1}{16\pi^2} \left (
\frac{1}{\epsilon}+\log(\omega^2/\mu^2) \right ).
\label{I0}
\end{eqnarray}
We also introduce
\begin{eqnarray}
\bar I (q^2,\omega) \equiv I(q^2,\omega) - I(0,\omega).
\end{eqnarray} 
In the Feynman parametric form we equivalently have
\begin{eqnarray}
I(q^2 , \omega ) &=& -\frac1{(4\pi)^2 } \int d\omega \rho(\omega) \nonumber \\
&\times& \int_0^1 dx \log\left[\omega^2 +x(1-x) q^2 \right].
\label{eq:ifeynman}
\end{eqnarray} 
The imaginary part yields
\begin{eqnarray}
\frac1\pi  {\rm Im} I (q^2,\omega) = \frac1{16\pi^2}
\sqrt{1-\frac{\omega^2}{4 q^2}} \theta ( q^2 - 4\omega^2 ).  
\end{eqnarray} 
Thus the once-subtracted dispersion relation,
\begin{eqnarray}
\bar I (q^2,\omega) =
\frac{q^2} {\pi} \int_{4w^2}^\infty \frac{dt}{t} \, \frac{{\rm Im} I
(t,\omega) }{t-q^2-i0^+},
\end{eqnarray} 
holds. The asymptotic behavior for large Euclidean $-q^2$ is
\begin{eqnarray}
&& \!\!\!\!\! I(q^2,\omega) = \frac1{16\pi^2} \{ 2-\frac{1}{\epsilon}-\log (-q^2/\mu^2) +\label{Ihighq} \\
&& \!\!\!\!\! \frac{ 2 \omega^2}{q^2} \left[ \log (-q^2/\omega^2) + 1 \right] + 
\frac{ 2 \omega^4}{q^4} \left[ \log (-q^2/\omega^2) - \frac{1}{2} \right] \dots \}. 
\nonumber
\end{eqnarray} 
At low $q^2$  we have 
\begin{eqnarray}
\bar I(q^2,\omega) = \frac1{16\pi^2} \left\{
\frac{q^2}{6\omega^2}+\frac{q^4}{60\omega^4}+  \frac{q^6}{420\omega^6}+    
 \dots \right\}.
\end{eqnarray} 

\subsection{Three-point integral}

The three-point one-loop integral is defined as
\begin{eqnarray}
&& K((q_1-q_2)^2,q_1^2,q_2^2,\omega) = \frac1 i \int {d^4 k \over (2\pi)^4} {1\over k^2
- \omega^2 + i 0^+} \times \nonumber \\
&& {1\over (k-q_1)^2 - \omega^2 + i 0^+} {1\over (k-q_2)^2 - \omega^2 + i 0^+} . \nonumber \\
\end{eqnarray} 
We analyze it with the dimensional regularization, 
and for the case where the virtuality 
of one of the external line vanishes, $(q_1-q_2)^2=0$ (massless pion). 
We immediately find the result 
\begin{equation}
K(0,0,0,\omega)=-\frac{1}{16 \pi^2} \frac{1}{2\omega ^{2}}.  \label{c0000}
\end{equation}

The following Feynman parameterization is useful:
\begin{equation}
\frac{1}{abc}=2\int_{0}^{1}dx\int_{0}^{1}dy\frac{x}{[xya+x(1-y)b+(1-x)c]^{3}}%
.  \label{Feynman}
\end{equation}
Carrying the momentum integration and introducing $z=2y-1$ yields 
\begin{eqnarray}
&&K(0,\frac{1+A}{2}q^{2},\frac{1-A}{2}q^{2},\omega)=
\frac{1}{16 \pi^2} \int_{0}^{1}dx\int_{-1}^{1}dz \times \nonumber \\
&& \frac{x}{(1-x)x(1+Az)q^{2}-2\omega ^{2}}.  \label{c0one}
\end{eqnarray}
At $A=0$ we find 
\begin{eqnarray}
&&K(0,\frac{1}{2}q^{2},\frac{1}{2}q^{2},\omega)=\label{c0A0} \\
&& \frac{1}{16 \pi^2} \frac{2}{q^{2}\sqrt{1-8\omega
^{2}/q^{2}}}\log \frac{1+\sqrt{1-8w^{2}/q^{2}}}{1-\sqrt{1-8w^{2}/q^{2}}}. \nonumber
\end{eqnarray}
At low $q^{2}$ the expansion is 
\begin{eqnarray}
&& K(0,\frac{1}{2}q^{2},\frac{1}{2}q^{2},\omega)=  \label{c0A0lowq}\\
&& -\frac{1}{16 \pi^2} \left [ \frac{1}{2\omega ^{2}}+\frac{%
q^{2}}{24\omega ^{4}}+\frac{q^{4}}{240\omega ^{4}}+... \right ] , \nonumber
\end{eqnarray}
and an high $q^{2}$%
\begin{eqnarray}
&& K(0,\frac{1}{2}q^{2},\frac{1}{2}q^{2})=
 -\frac{1}{16 \pi^2} \left [ 2\log \left( -2\omega
^{2}/q^{2}\right) \frac{1}{q^{2}}
\right . \nonumber \\
&& \left . -8\left[ \log \left( -2\omega
^{2}/q^{2}\right) +1\right] \frac{\omega ^{2}}{q^{4}}+... \right ] .  \label{c0A0highq}
\end{eqnarray}
The integral over $x$ in Eq.~(\ref{c0one}) yields 
\begin{eqnarray}
&& K(0,\frac{1+A}{2}q^{2},\frac{1-A}{2}q^{2},\omega) = \label{c0A} \\
&& \frac{1}{16 \pi^2} \int_{-1}^{1}dz\frac{1}{%
(1+Az)q^{2}s}\log \frac{1+s}{1-s},  \nonumber \\
&& s =\sqrt{1-8\omega ^{2}/(q^{2}(1+Az))}. \nonumber
\end{eqnarray}
At low $q^2$ we have 
\begin{eqnarray}
&& K(0,\frac{1+A}{2}q^{2},\frac{1-A}{2}q^{2},\omega)=  \label{c0Alowq}
\\
&& -\frac{1}{16 \pi^2} \left [ \frac{1}{2\omega ^{2}}+\frac{%
q^{2}}{24\omega ^{4}}+\frac{(A^2+3)q^{4}}{720\omega ^{4}}+... \right ] , \nonumber
\end{eqnarray}
The large-$q^{2}$ expansion  produces 
\begin{eqnarray}
&& K(0,\frac{1+A}{2}q^{2},\frac{1-A}{2}q^{2},\omega) \label{newhq}\\
&& = -\frac{1}{16 \pi^2} \int_{-1}^{1}dz\left\{ \frac{1}{q^{2}(1+Az)}
\log \left( -\frac{2\omega
^{2}}{q^{2}(1+Az)}\right) \right . \nonumber \\ 
&& \left . +\frac{4\omega ^{2}}{q^{4}(1+Az)^2}\left[ \log
\left( -\frac{2\omega ^{2}}{q^{2}(1+Az)}\right) +1\right] +....\right\} 
\nonumber
\end{eqnarray}
The integral over $z$ in Eq.~(\ref{c0A}) gives finally the simple
general result 
\begin{eqnarray}
&& K(0,\frac{1+A}{2}q^{2},\frac{1-A}{2}q^{2}) \label{c0Afull} \\
&& =\frac{1}{32 \pi^2 Aq^{2}}\left[
\left( \log \frac{1+s_{+}}{1-s_{+}}\right) ^{2}-\left( \log \frac{1+s_{-}}{%
1-s_{-}}\right) ^{2}\right] ,  \nonumber \\
s_{\pm } &=&\sqrt{1-8\omega ^{2}/(q^{2}(1\pm A))}. \nonumber 
\end{eqnarray}

\section{Discontinuity in the Bjorken limit} \label{sec:appb}

Let us consider the one-loop function
\begin{eqnarray} 
T(p,q) &=& i \int {d^4 k \over (2\pi)^4} {1\over [(k-p)^2 - \omega^2 + i
0^+]^2} \times \\ 
&& {1\over k^2 - \omega^2 + i 0^+} {1\over (q-k-p)^2 - \omega^2 + i 0^+}. \nonumber 
\end{eqnarray} 
The discontinuity in the $s=(p+q)^2 $ channel may be computed through the
Cutkosky rules,
\begin{eqnarray} 
&& {\rm Disc} \, T(p,q) = i \int {d^4 k \over (2\pi)^4} {1\over [(k-p)^2 -
\omega^2 + i 0^+]^2} \nonumber \\ && (-2\pi i )^2 \delta^+( k^2 - \omega^2 ) \delta^+ [
(q-k-p)^2 - \omega^2 ].
\end{eqnarray} 
We choose the reference frame of the target at rest 
\begin{eqnarray}
p=(m, \vec{0}_\perp , 0 ), \; q=(q_0 , \vec{0}_\perp , q_3 ), \;
q^2 = -Q^2 = q_0^2 - \vec q^2  . \nonumber \\
\end{eqnarray} 
One gets then 
\begin{eqnarray}
q_0 &=& \frac{Q^2}{2m_\pi x}, \\ 
q_3 &=&  \frac{Q^2}{2m_\pi  x}\sqrt{1+
\frac{4m_\pi^2 x^2}{Q^2}} \to \frac{Q^2}{2m_\pi  x} + m_\pi  x + \dots \nonumber 
\end{eqnarray} 

In the light-cone coordinates, defined as
\begin{eqnarray}
k^+ &=& k^0 + k^3, \; k^- = k^0 - k^3, \; \vec k_\perp =(k^1 ,
k^2 ), \nonumber \\ dk^0 dk^3 &=& \frac12 dk^+ dk^- ,
\end{eqnarray} 
one obtains 
\begin{eqnarray}
 q^+ &=& q^0 + q^3 \to
\frac{Q^2}{m_\pi  x}, \nonumber \\ q^- &=& q^0 - q^3 \to - m_\pi  x,
\end{eqnarray} 
and also  
\begin{eqnarray}
&&\delta^+ \left[ (k-p-q)^2 -\omega^2 \right] \to \frac{m_\pi  x}{Q^2} 
\delta \left[ k^- -(1-x) m_\pi  \right], \nonumber \\ 
&&\delta^+ \left[ k^2 -\omega^2 \right] \to \frac{1}{m_\pi (1-x)}
\delta \left[ k^+ -\frac{\vec k_\perp^2 +\omega^2 }{m_\pi (1-x)} \right] . \nonumber \\
\end{eqnarray} 
Thus finally we get 
\begin{eqnarray}
&& {\rm Im} T(p,q)  \to \\
&& \frac{ x(1-x)}{Q^2} \int \frac
{d^2 \vec k_\perp}{(2\pi)^2 } \frac1{\left[ \vec k_\perp^2 + \omega^2 - m_\pi^2
x (1-x) ] \right]^2 } . \nonumber
\end{eqnarray} 

%\bibliographystyle{prsty}

%\bibliography{spec}

\begin{thebibliography}{66}
\expandafter\ifx\csname natexlab\endcsname\relax\def\natexlab#1{#1}\fi
\expandafter\ifx\csname bibnamefont\endcsname\relax
  \def\bibnamefont#1{#1}\fi
\expandafter\ifx\csname bibfnamefont\endcsname\relax
  \def\bibfnamefont#1{#1}\fi
\expandafter\ifx\csname citenamefont\endcsname\relax
  \def\citenamefont#1{#1}\fi
\expandafter\ifx\csname url\endcsname\relax
  \def\url#1{\texttt{#1}}\fi
\expandafter\ifx\csname urlprefix\endcsname\relax\def\urlprefix{URL }\fi
\providecommand{\bibinfo}[2]{#2}
\providecommand{\eprint}[2][]{\url{#2}}

\bibitem[{\citenamefont{Vogl and Weise}(1991)}]{NJLrev:reg}
\bibinfo{author}{\bibfnamefont{U.}~\bibnamefont{Vogl}} \bibnamefont{and}
  \bibinfo{author}{\bibfnamefont{W.}~\bibnamefont{Weise}},
  \bibinfo{journal}{Prog. Part. Nucl. Phys.} \textbf{\bibinfo{volume}{27}},
  \bibinfo{pages}{195} (\bibinfo{year}{1991}).

\bibitem[{\citenamefont{Klevansky}(1992)}]{NJLrev:klev}
\bibinfo{author}{\bibfnamefont{S.~P.} \bibnamefont{Klevansky}},
  \bibinfo{journal}{Rev. Mod. Phys.} \textbf{\bibinfo{volume}{64}},
  \bibinfo{pages}{649} (\bibinfo{year}{1992}).

\bibitem[{\citenamefont{Volkov}(1993)}]{NJLrev:volkov}
\bibinfo{author}{\bibfnamefont{M.~K.} \bibnamefont{Volkov}},
  \bibinfo{journal}{Part. and Nuclei} \textbf{\bibinfo{volume}{B24}},
  \bibinfo{pages}{1} (\bibinfo{year}{1993}).

\bibitem[{\citenamefont{Hatsuda and Kunihiro}(1994)}]{NJLrev:jap}
\bibinfo{author}{\bibfnamefont{T.}~\bibnamefont{Hatsuda}} \bibnamefont{and}
  \bibinfo{author}{\bibfnamefont{T.}~\bibnamefont{Kunihiro}},
  \bibinfo{journal}{Phys. Rep.} \textbf{\bibinfo{volume}{247}},
  \bibinfo{pages}{221} (\bibinfo{year}{1994}).

\bibitem[{\citenamefont{Christov et~al.}(1996)\citenamefont{Christov, Blotz,
  Kim, Pobylitsa, Watabe, Meissner, Arriola, and Goeke}}]{NJLrev:Bo}
\bibinfo{author}{\bibfnamefont{C.~V.} \bibnamefont{Christov}},
  \bibinfo{author}{\bibfnamefont{A.}~\bibnamefont{Blotz}},
  \bibinfo{author}{\bibfnamefont{H.-C.} \bibnamefont{Kim}},
  \bibinfo{author}{\bibfnamefont{P.}~\bibnamefont{Pobylitsa}},
  \bibinfo{author}{\bibfnamefont{T.}~\bibnamefont{Watabe}},
  \bibinfo{author}{\bibfnamefont{T.}~\bibnamefont{Meissner}},
  \bibinfo{author}{\bibfnamefont{E. Ruiz} \bibnamefont{Arriola}},
  \bibnamefont{and} \bibinfo{author}{\bibfnamefont{K.}~\bibnamefont{Goeke}},
  \bibinfo{journal}{Prog. Part. Nucl. Phys.} \textbf{\bibinfo{volume}{37}},
  \bibinfo{pages}{91} (\bibinfo{year}{1996}).

\bibitem[{\citenamefont{Alkofer et~al.}(1996)\citenamefont{Alkofer, Reinhardt,
  and Weigel}}]{NJLrev:Tue}
\bibinfo{author}{\bibfnamefont{R.}~\bibnamefont{Alkofer}},
  \bibinfo{author}{\bibfnamefont{H.}~\bibnamefont{Reinhardt}},
  \bibnamefont{and} \bibinfo{author}{\bibfnamefont{H.}~\bibnamefont{Weigel}},
  \bibinfo{journal}{Phys. Rep.} \textbf{\bibinfo{volume}{265}},
  \bibinfo{pages}{139} (\bibinfo{year}{1996}).

\bibitem[{\citenamefont{Ripka}(1997)}]{NJLrev:ripka}
\bibinfo{author}{\bibfnamefont{G.}~\bibnamefont{Ripka}},
  \emph{\bibinfo{title}{Quarks Bound by Chiral Fields}}
  (\bibinfo{publisher}{Clarendon Press}, \bibinfo{address}{Oxford},
  \bibinfo{year}{1997}).

\bibitem[{\citenamefont{Ruiz Arriola}(2002)}]{Ru02}
\bibinfo{author}{\bibfnamefont{E. Ruiz} \bibnamefont{Arriola}},
  \bibinfo{journal}{Acta Phys. Pol.} \textbf{\bibinfo{volume}{B33}},
  \bibinfo{pages}{4443} (\bibinfo{year}{2002}).

\bibitem[{\citenamefont{Itzykson and Zuber}(1980)}]{IZ80}
\bibinfo{author}{\bibfnamefont{C.}~\bibnamefont{Itzykson}} \bibnamefont{and}
  \bibinfo{author}{\bibfnamefont{J.~B.} \bibnamefont{Zuber}},
  \emph{\bibinfo{title}{Quantum Field Theory}}
  (\bibinfo{publisher}{McGraw-Hill}, \bibinfo{address}{New York},
  \bibinfo{year}{1980}).

\bibitem[{\citenamefont{Delbourgo and West}(1977)}]{DW77}
\bibinfo{author}{\bibfnamefont{R.}~\bibnamefont{Delbourgo}} \bibnamefont{and}
  \bibinfo{author}{\bibfnamefont{P.~C.} \bibnamefont{West}},
  \bibinfo{journal}{J. Phys.} \textbf{\bibinfo{volume}{A10}},
  \bibinfo{pages}{1049} (\bibinfo{year}{1977}).

\bibitem[{\citenamefont{Delbourgo}(1979)}]{De79}
\bibinfo{author}{\bibfnamefont{R.}~\bibnamefont{Delbourgo}},
  \bibinfo{journal}{Nuovo Cimento} \textbf{\bibinfo{volume}{A49}}
  (\bibinfo{year}{1979}).

\bibitem[{\citenamefont{Arriola}(2001)}]{Ru01}
\bibinfo{author}{\bibfnamefont{E. Ruiz} \bibnamefont{Arriola}}, in
  \emph{\bibinfo{booktitle}{proceedings of the Workshop on Lepton Scattering,
  Hadrons and QCD, Adelaide, Australia, 2001}}, edited by
  \bibinfo{editor}{\bibfnamefont{W.}~\bibnamefont{Melnitchouk}},
  \bibinfo{editor}{\bibfnamefont{A.}~\bibnamefont{Schreiber}},
  \bibinfo{editor}{\bibfnamefont{P.}~\bibnamefont{Tandy}}, \bibnamefont{and}
  \bibinfo{editor}{\bibfnamefont{A.~W.} \bibnamefont{Thomas}}
  (\bibinfo{publisher}{World Scientific}, \bibinfo{address}{Singapore},
  \bibinfo{year}{2001}), \bibinfo{note}{hep-ph/0107087}.

\bibitem[{\citenamefont{Jaffe and Ross}(1980)}]{JR80}
\bibinfo{author}{\bibfnamefont{R.~L.} \bibnamefont{Jaffe}} \bibnamefont{and}
  \bibinfo{author}{\bibfnamefont{G.~C.} \bibnamefont{Ross}},
  \bibinfo{journal}{Phys. Lett.} \textbf{\bibinfo{volume}{B93}},
  \bibinfo{pages}{313} (\bibinfo{year}{1980}).

\bibitem[{\citenamefont{Jaffe}(1986)}]{Ja85}
\bibinfo{author}{\bibfnamefont{R.~L.} \bibnamefont{Jaffe}}, in
  \emph{\bibinfo{booktitle}{proceedings of the Los Alamos School, 1985}},
  edited by \bibinfo{editor}{\bibfnamefont{M.~B.} \bibnamefont{Johnson}}
  \bibnamefont{and}
  \bibinfo{editor}{\bibfnamefont{A.}~\bibnamefont{Picklesimer}}
  (\bibinfo{publisher}{Wiley}, \bibinfo{address}{New York},
  \bibinfo{year}{1986}).

\bibitem[{\citenamefont{Davidson and Arriola}(1995)}]{DR95}
\bibinfo{author}{\bibfnamefont{R.~M.} \bibnamefont{Davidson}} \bibnamefont{and}
  \bibinfo{author}{\bibfnamefont{E. Ruiz} \bibnamefont{Arriola}},
  \bibinfo{journal}{Phys. Lett.} \textbf{\bibinfo{volume}{B359}},
  \bibinfo{pages}{273} (\bibinfo{year}{1995}).

\bibitem[{\citenamefont{Davidson and Arriola}(2002)}]{DR02}
\bibinfo{author}{\bibfnamefont{R.~M.} \bibnamefont{Davidson}} \bibnamefont{and}
  \bibinfo{author}{\bibfnamefont{E. Ruiz} \bibnamefont{Arriola}},
  \bibinfo{journal}{Act. Phys. Pol.} \textbf{\bibinfo{volume}{B33}},
  \bibinfo{pages}{1791} (\bibinfo{year}{2002}).

\bibitem[{\citenamefont{Arriola and Broniowski}(2002)}]{RB02}
\bibinfo{author}{\bibfnamefont{E. Ruiz} \bibnamefont{Arriola}} \bibnamefont{and}
  \bibinfo{author}{\bibfnamefont{W.}~\bibnamefont{Broniowski}},
  \bibinfo{journal}{Phys. Rev} \textbf{\bibinfo{volume}{D66}},
  \bibinfo{pages}{094016} (\bibinfo{year}{2002}).

\bibitem[{\citenamefont{Bowman et~al.}(2002{\natexlab{a}})\citenamefont{Bowman,
  Heller, and Williams}}]{Bowman:2002bm}
\bibinfo{author}{\bibfnamefont{P.~O.} \bibnamefont{Bowman}},
  \bibinfo{author}{\bibfnamefont{U.~M.} \bibnamefont{Heller}},
  \bibnamefont{and} \bibinfo{author}{\bibfnamefont{A.~G.}
  \bibnamefont{Williams}}, \bibinfo{journal}{Phys. Rev.}
  \textbf{\bibinfo{volume}{D66}}, \bibinfo{pages}{014505}
  (\bibinfo{year}{2002}{\natexlab{a}}).

\bibitem[{\citenamefont{Bowman et~al.}(2002{\natexlab{b}})\citenamefont{Bowman,
  Heller, Leinweber, and Williams}}]{Bowman:2002kn}
\bibinfo{author}{\bibfnamefont{P.~O.} \bibnamefont{Bowman}},
  \bibinfo{author}{\bibfnamefont{U.~M.} \bibnamefont{Heller}},
  \bibinfo{author}{\bibfnamefont{D.~B.} \bibnamefont{Leinweber}},
  \bibnamefont{and} \bibinfo{author}{\bibfnamefont{A.~G.}
  \bibnamefont{Williams}} (\bibinfo{year}{2002}{\natexlab{b}}),
  \eprint{hep-lat/0209129}.

\bibitem[{\citenamefont{Haeri}(1988)}]{Ha88}
\bibinfo{author}{\bibfnamefont{B.}~\bibnamefont{Haeri}},
  \bibinfo{journal}{Phys. Rev.} \textbf{\bibinfo{volume}{D38}},
  \bibinfo{pages}{3799} (\bibinfo{year}{1988}).

\bibitem[{\citenamefont{Lee and Wick}(1969)}]{LW69}
\bibinfo{author}{\bibfnamefont{T.~D.} \bibnamefont{Lee}} \bibnamefont{and}
  \bibinfo{author}{\bibfnamefont{G.~C.} \bibnamefont{Wick}},
  \bibinfo{journal}{Nucl. Phys.} \textbf{\bibinfo{volume}{B9}},
  \bibinfo{pages}{209} (\bibinfo{year}{1969}).

\bibitem[{\citenamefont{Cutkosky et~al.}(1969)\citenamefont{Cutkosky,
  Landshoff, Olive, and Polkinghorne}}]{CL69}
\bibinfo{author}{\bibfnamefont{R.~E.} \bibnamefont{Cutkosky}},
  \bibinfo{author}{\bibfnamefont{P.~V.} \bibnamefont{Landshoff}},
  \bibinfo{author}{\bibfnamefont{D.~I.} \bibnamefont{Olive}}, \bibnamefont{and}
  \bibinfo{author}{\bibfnamefont{J.}~\bibnamefont{Polkinghorne}},
  \bibinfo{journal}{Nucl. Phys.} \textbf{\bibinfo{volume}{B12}},
  \bibinfo{pages}{281} (\bibinfo{year}{1969}).

\bibitem[{\citenamefont{Boulware and Gross}(1984)}]{BG84}
\bibinfo{author}{\bibfnamefont{D.~G.} \bibnamefont{Boulware}} \bibnamefont{and}
  \bibinfo{author}{\bibfnamefont{D.~J.} \bibnamefont{Gross}},
  \bibinfo{journal}{Nucl. Phys.} \textbf{\bibinfo{volume}{B233}},
  \bibinfo{pages}{1} (\bibinfo{year}{1984}).

\bibitem[{\citenamefont{Prasza\l{}owicz and Rostworowski}(2001)}]{PR01}
\bibinfo{author}{\bibfnamefont{M.}~\bibnamefont{Prasza\l{}owicz}}
  \bibnamefont{and}
  \bibinfo{author}{\bibfnamefont{A.}~\bibnamefont{Rostworowski}},
  \bibinfo{journal}{Phys. Rev.} \textbf{\bibinfo{volume}{D64}},
  \bibinfo{pages}{074003} (\bibinfo{year}{2001}).

\bibitem[{\citenamefont{Ioffe}(2002)}]{Io02}
\bibinfo{author}{\bibfnamefont{B.~L.} \bibnamefont{Ioffe}},
  \bibinfo{type}{Tech. Rep.} (\bibinfo{year}{2002}), \eprint{hep-ph/0207191}.

\bibitem[{\citenamefont{Mikhailov and Radyushkin}(1989)}]{Mikhailov:1989nz}
\bibinfo{author}{\bibfnamefont{S.~V.} \bibnamefont{Mikhailov}}
  \bibnamefont{and} \bibinfo{author}{\bibfnamefont{A.~V.}
  \bibnamefont{Radyushkin}}, \bibinfo{journal}{Sov. J. Nucl. Phys.}
  \textbf{\bibinfo{volume}{49}}, \bibinfo{pages}{494} (\bibinfo{year}{1989}).

\bibitem[{\citenamefont{Mikhailov and Radyushkin}(1992)}]{Mikhailov:1992}
\bibinfo{author}{\bibfnamefont{S.~V.} \bibnamefont{Mikhailov}}
  \bibnamefont{and} \bibinfo{author}{\bibfnamefont{A.~V.}
  \bibnamefont{Radyushkin}}, \bibinfo{journal}{Phys. Rev.}
  \textbf{\bibinfo{volume}{D45}}, \bibinfo{pages}{1754} (\bibinfo{year}{1992}).

\bibitem[{\citenamefont{Bakulev and Mikhailov}(2002)}]{Bakulev:2002hk}
\bibinfo{author}{\bibfnamefont{A.~P.} \bibnamefont{Bakulev}} \bibnamefont{and}
  \bibinfo{author}{\bibfnamefont{S.~V.} \bibnamefont{Mikhailov}},
  \bibinfo{journal}{Phys. Rev.} \textbf{\bibinfo{volume}{D65}},
  \bibinfo{pages}{114511} (\bibinfo{year}{2002}).

\bibitem[{\citenamefont{Dorokhov and Broniowski}(2002)}]{Dorokhov:2001wx}
\bibinfo{author}{\bibfnamefont{A.~E.} \bibnamefont{Dorokhov}} \bibnamefont{and}
  \bibinfo{author}{\bibfnamefont{W.}~\bibnamefont{Broniowski}},
  \bibinfo{journal}{Phys. Rev.} \textbf{\bibinfo{volume}{D65}},
  \bibinfo{pages}{094007} (\bibinfo{year}{2002}).

\bibitem[{\citenamefont{Belyaev and Ioffe}(1982)}]{BI82}
\bibinfo{author}{\bibfnamefont{V.~M.} \bibnamefont{Belyaev}} \bibnamefont{and}
  \bibinfo{author}{\bibfnamefont{B.~L.} \bibnamefont{Ioffe}},
  \bibinfo{journal}{Sov. Phys. JETP} \textbf{\bibinfo{volume}{56}},
  \bibinfo{pages}{493} (\bibinfo{year}{1982}).

\bibitem[{\citenamefont{Ioffe and Zyablyuk}(2002)}]{IoZa}
\bibinfo{author}{\bibfnamefont{B.~L.} \bibnamefont{Ioffe}} \bibnamefont{and}
  \bibinfo{author}{\bibfnamefont{K.~N.} \bibnamefont{Zyablyuk}},
  \bibinfo{type}{Tech. Rep.} (\bibinfo{year}{2002}), \eprint{hep-ph/0207183}.

\bibitem[{\citenamefont{Renner}(1968)}]{Re67}
\bibinfo{author}{\bibfnamefont{B.}~\bibnamefont{Renner}},
  \emph{\bibinfo{title}{Current Algebras and Their Applications}}
  (\bibinfo{publisher}{Pergamon Press}, \bibinfo{address}{New York},
  \bibinfo{year}{1968}).

\bibitem[{\citenamefont{Delbourgo}(1999)}]{De99}
\bibinfo{author}{\bibfnamefont{R.}~\bibnamefont{Delbourgo}},
  \bibinfo{journal}{Austral. J. Phys.} \textbf{\bibinfo{volume}{52}},
  \bibinfo{pages}{681} (\bibinfo{year}{1999}).

\bibitem[{\citenamefont{Sauli}(2002)}]{Sa02}
\bibinfo{author}{\bibfnamefont{V.}~\bibnamefont{Sauli}}, \bibinfo{type}{Tech.
  Rep.} (\bibinfo{year}{2002}), \eprint{hep-ph/0209046}.

\bibitem[{\citenamefont{Broniowski et~al.}(1996)\citenamefont{Broniowski,
  Ripka, Nikolov, and Goeke}}]{analyt}
\bibinfo{author}{\bibfnamefont{W.}~\bibnamefont{Broniowski}},
  \bibinfo{author}{\bibfnamefont{G.}~\bibnamefont{Ripka}},
  \bibinfo{author}{\bibfnamefont{E.~N.} \bibnamefont{Nikolov}},
  \bibnamefont{and} \bibinfo{author}{\bibfnamefont{K.}~\bibnamefont{Goeke}},
  \bibinfo{journal}{Zeit. Phys. A} \textbf{\bibinfo{volume}{354}},
  \bibinfo{pages}{421} (\bibinfo{year}{1996}).

\bibitem[{\citenamefont{Pagels and Stokar}(1979)}]{PS79}
\bibinfo{author}{\bibfnamefont{H.}~\bibnamefont{Pagels}} \bibnamefont{and}
  \bibinfo{author}{\bibfnamefont{S.}~\bibnamefont{Stokar}},
  \bibinfo{journal}{Phys. Rev.} \textbf{\bibinfo{volume}{D20}},
  \bibinfo{pages}{2947} (\bibinfo{year}{1979}).

\bibitem[{\citenamefont{Weinberg}(1967)}]{We67}
\bibinfo{author}{\bibfnamefont{S.}~\bibnamefont{Weinberg}},
  \bibinfo{journal}{Phys. Rev. Lett.} \textbf{\bibinfo{volume}{18}},
  \bibinfo{pages}{507} (\bibinfo{year}{1967}).

\bibitem[{\citenamefont{C.~Bernard and Weinberg}(1975)}]{BD75}
\bibinfo{author}{\bibfnamefont{J.~L.} \bibnamefont{C.~Bernard},
  \bibfnamefont{A.~Duncan}} \bibnamefont{and}
  \bibinfo{author}{\bibfnamefont{S.}~\bibnamefont{Weinberg}},
  \bibinfo{journal}{Phys. Rev.} \textbf{\bibinfo{volume}{D12}},
  \bibinfo{pages}{792} (\bibinfo{year}{1975}).

\bibitem[{\citenamefont{Broniowski}(1999)}]{Broniowski:1999dm}
\bibinfo{author}{\bibfnamefont{W.}~\bibnamefont{Broniowski}}
  (\bibinfo{year}{1999}), \eprint{hep-ph/9911204}.

\bibitem[{\citenamefont{Tarrach}(1979)}]{Tarrach:1979ta}
\bibinfo{author}{\bibfnamefont{R.}~\bibnamefont{Tarrach}}, \bibinfo{journal}{Z.
  Phys.} \textbf{\bibinfo{volume}{C2}}, \bibinfo{pages}{221}
  (\bibinfo{year}{1979}).

\bibitem[{\citenamefont{Faccioli et~al.}(2002)\citenamefont{Faccioli, Schwenk,
  and Shuryak}}]{Faccioli:2002jd}
\bibinfo{author}{\bibfnamefont{P.}~\bibnamefont{Faccioli}},
  \bibinfo{author}{\bibfnamefont{A.}~\bibnamefont{Schwenk}}, \bibnamefont{and}
  \bibinfo{author}{\bibfnamefont{E.~V.} \bibnamefont{Shuryak}}
  (\bibinfo{year}{2002}), \eprint{hep-ph/0202027}.

\bibitem[{\citenamefont{et~al.}(1976)}]{Bebek}
\bibinfo{author}{\bibfnamefont{C.~J.~B.} \bibnamefont{et~al.}},
  \bibinfo{journal}{Phys. Rev. Lett} \textbf{\bibinfo{volume}{37}},
  \bibinfo{pages}{1693} (\bibinfo{year}{1976}).

\bibitem[{\citenamefont{Volmer et~al.}(2001)}]{Volmer:2000ek}
\bibinfo{author}{\bibfnamefont{J.}~\bibnamefont{Volmer}} \bibnamefont{et~al.}
  (\bibinfo{collaboration}{The Jefferson Lab F(pi)}), \bibinfo{journal}{Phys.
  Rev. Lett.} \textbf{\bibinfo{volume}{86}}, \bibinfo{pages}{1713}
  (\bibinfo{year}{2001}), \eprint{nucl-ex/0010009}.

\bibitem[{\citenamefont{Blok et~al.}(2002)\citenamefont{Blok, Huber, and
  Mack}}]{Blok}
\bibinfo{author}{\bibfnamefont{H.~P.} \bibnamefont{Blok}},
  \bibinfo{author}{\bibfnamefont{G.~M.} \bibnamefont{Huber}}, \bibnamefont{and}
  \bibinfo{author}{\bibfnamefont{D.~J.} \bibnamefont{Mack}}
  (\bibinfo{year}{2002}), \bibinfo{note}{nucl-ex/0208011}.

\bibitem[{\citenamefont{Amendolia et~al.}(1986)}]{Amendolia:1986wj}
\bibinfo{author}{\bibfnamefont{S.~R.} \bibnamefont{Amendolia}}
  \bibnamefont{et~al.} (\bibinfo{collaboration}{NA7}), \bibinfo{journal}{Nucl.
  Phys.} \textbf{\bibinfo{volume}{B277}}, \bibinfo{pages}{168}
  (\bibinfo{year}{1986}).

\bibitem[{\citenamefont{Lepage and Brodsky}(1980)}]{BL80}
\bibinfo{author}{\bibfnamefont{G.~P.} \bibnamefont{Lepage}} \bibnamefont{and}
  \bibinfo{author}{\bibfnamefont{S.~J.} \bibnamefont{Brodsky}},
  \bibinfo{journal}{Phys. Rev.} \textbf{\bibinfo{volume}{D22}},
  \bibinfo{pages}{2157} (\bibinfo{year}{1980}).

\bibitem[{\citenamefont{Dorokhov}(2002{\natexlab{a}})}]{Do02}
\bibinfo{author}{\bibfnamefont{A.~E.} \bibnamefont{Dorokhov}}
  (\bibinfo{year}{2002}{\natexlab{a}}), \bibinfo{note}{talk presented at the
  37th Rencontres de Moriond on QCD and Hadronic Interactions, Les Arcs,
  France, 16-23 March 2002, hep-ph/0206088}.

\bibitem[{\citenamefont{Dorokhov}(2002{\natexlab{b}})}]{Do02b}
\bibinfo{author}{\bibfnamefont{A.~E.} \bibnamefont{Dorokhov}},
  \bibinfo{journal}{Pis'ma ZhETF} \textbf{\bibinfo{volume}{77}},
  \bibinfo{pages}{68} (\bibinfo{year}{2002}{\natexlab{b}}).

\bibitem[{\citenamefont{{M\"uller}}(1995)}]{Mu95}
\bibinfo{author}{\bibfnamefont{D.}~\bibnamefont{{M\"uller}}},
  \bibinfo{journal}{Phys. Rev.} \textbf{\bibinfo{volume}{D51}},
  \bibinfo{pages}{3855} (\bibinfo{year}{1995}).

\bibitem[{\citenamefont{Schmedding and Yakovlev}(2000)}]{SY00}
\bibinfo{author}{\bibfnamefont{A.}~\bibnamefont{Schmedding}} \bibnamefont{and}
  \bibinfo{author}{\bibfnamefont{O.}~\bibnamefont{Yakovlev}},
  \bibinfo{journal}{Phys. Rev.} \textbf{\bibinfo{volume}{D62}},
  \bibinfo{pages}{116002} (\bibinfo{year}{2000}).

\bibitem[{\citenamefont{Bakulev et~al.}(2002)\citenamefont{Bakulev, Mikhailov,
  and Stefanis}}]{Bakulev:2002uc}
\bibinfo{author}{\bibfnamefont{A.~P.} \bibnamefont{Bakulev}},
  \bibinfo{author}{\bibfnamefont{S.~V.} \bibnamefont{Mikhailov}},
  \bibnamefont{and} \bibinfo{author}{\bibfnamefont{N.~G.}
  \bibnamefont{Stefanis}} (\bibinfo{year}{2002}), \eprint{hep-ph/0212250}.

\bibitem[{\citenamefont{{CLEO Collaboration (J. Gronberg et
  al.)}}(1998)}]{CLEO98}
\bibinfo{author}{\bibnamefont{{CLEO Collaboration (J. Gronberg et al.)}}},
  \bibinfo{journal}{Phys. Rev.} \textbf{\bibinfo{volume}{D57}},
  \bibinfo{pages}{33} (\bibinfo{year}{1998}).

\bibitem[{\citenamefont{Adler}(1971)}]{adler}
\bibinfo{author}{\bibfnamefont{S.~L.} \bibnamefont{Adler}},
  \bibinfo{journal}{Phys. Rev.} \textbf{\bibinfo{volume}{D4}},
  \bibinfo{pages}{3497} (\bibinfo{year}{1971}).

\bibitem[{\citenamefont{Terent'ev}(1972)}]{terentev}
\bibinfo{author}{\bibfnamefont{M.~V.} \bibnamefont{Terent'ev}},
  \bibinfo{journal}{Phys. Lett.} \textbf{\bibinfo{volume}{38B}},
  \bibinfo{pages}{419} (\bibinfo{year}{1972}).

\bibitem[{\citenamefont{Aviv and Zee}(1972)}]{aviv}
\bibinfo{author}{\bibfnamefont{R.}~\bibnamefont{Aviv}} \bibnamefont{and}
  \bibinfo{author}{\bibfnamefont{A.}~\bibnamefont{Zee}},
  \bibinfo{journal}{Phys. Rev.} \textbf{\bibinfo{volume}{D5}},
  \bibinfo{pages}{2372} (\bibinfo{year}{1972}).

\bibitem[{\citenamefont{Weigel et~al.}(1999)\citenamefont{Weigel, Arriola, and
  Gamberg}}]{WRG99}
\bibinfo{author}{\bibfnamefont{H.}~\bibnamefont{Weigel}},
  \bibinfo{author}{\bibfnamefont{E. Ruiz} \bibnamefont{Arriola}},
  \bibnamefont{and} \bibinfo{author}{\bibfnamefont{L.}~\bibnamefont{Gamberg}},
  \bibinfo{journal}{Nucl. Phys.} \textbf{\bibinfo{volume}{B560}},
  \bibinfo{pages}{383} (\bibinfo{year}{1999}).

\bibitem[{\citenamefont{Ellis}(1976)}]{Ellis:1976eh}
\bibinfo{author}{\bibfnamefont{J.~R.} \bibnamefont{Ellis}}
  (\bibinfo{year}{1976}), \bibinfo{note}{in {\em Les Houches 1976, Proceedings,
  Weak and Electromagnetic Interactions At High Energies}, Amsterdam 1977,
  1-114 and Preprint - ELLIS J (76,REC.JAN 77) 171p}.

\bibitem[{\citenamefont{Altarelli and Parisi}(1977)}]{AP77}
\bibinfo{author}{\bibfnamefont{G.}~\bibnamefont{Altarelli}} \bibnamefont{and}
  \bibinfo{author}{\bibfnamefont{G.}~\bibnamefont{Parisi}},
  \bibinfo{journal}{Nucl. Phys.} \textbf{\bibinfo{volume}{B126}},
  \bibinfo{pages}{298} (\bibinfo{year}{1977}).

\bibitem[{\citenamefont{Sutton et~al.}(1992)\citenamefont{Sutton, Martin,
  Roberts, and Stirling}}]{SMRS92}
\bibinfo{author}{\bibfnamefont{P.~J.} \bibnamefont{Sutton}},
  \bibinfo{author}{\bibfnamefont{A.~D.} \bibnamefont{Martin}},
  \bibinfo{author}{\bibfnamefont{R.~G.} \bibnamefont{Roberts}},
  \bibnamefont{and} \bibinfo{author}{\bibfnamefont{W.~J.}
  \bibnamefont{Stirling}}, \bibinfo{journal}{Phys. Rev.}
  \textbf{\bibinfo{volume}{D45}}, \bibinfo{pages}{2349} (\bibinfo{year}{1992}).

\bibitem[{\citenamefont{Gasser and Leutwyler}(1984)}]{GL84}
\bibinfo{author}{\bibfnamefont{J.}~\bibnamefont{Gasser}} \bibnamefont{and}
  \bibinfo{author}{\bibfnamefont{H.}~\bibnamefont{Leutwyler}},
  \bibinfo{journal}{Ann. Phys.} \textbf{\bibinfo{volume}{158}},
  \bibinfo{pages}{142} (\bibinfo{year}{1984}).

\bibitem[{\citenamefont{Gasser and Leutwyler}(1985)}]{GL85}
\bibinfo{author}{\bibfnamefont{J.}~\bibnamefont{Gasser}} \bibnamefont{and}
  \bibinfo{author}{\bibfnamefont{H.}~\bibnamefont{Leutwyler}},
  \bibinfo{journal}{Nucl. Phys.} \textbf{\bibinfo{volume}{B250}},
  \bibinfo{pages}{465} (\bibinfo{year}{1985}).

\bibitem[{\citenamefont{Chan}(1986)}]{Chan:1986jq}
\bibinfo{author}{\bibfnamefont{L.-H.} \bibnamefont{Chan}},
  \bibinfo{journal}{Phys. Rev. Lett.} \textbf{\bibinfo{volume}{57}},
  \bibinfo{pages}{1199} (\bibinfo{year}{1986}).

\bibitem[{\citenamefont{Arriola}(1991)}]{Ru91}
\bibinfo{author}{\bibfnamefont{E. Ruiz} \bibnamefont{Arriola}},
  \bibinfo{journal}{Phys. Lett.} \textbf{\bibinfo{volume}{B253}},
  \bibinfo{pages}{430} (\bibinfo{year}{1991}).

\bibitem[{\citenamefont{Schuren et~al.}(1992)\citenamefont{Schuren, Arriola,
  and Goeke}}]{SRG92}
\bibinfo{author}{\bibfnamefont{C.}~\bibnamefont{Schuren}},
  \bibinfo{author}{\bibfnamefont{E. Ruiz} \bibnamefont{Arriola}},
  \bibnamefont{and} \bibinfo{author}{\bibfnamefont{K.}~\bibnamefont{Goeke}},
  \bibinfo{journal}{Nucl. Phys.} \textbf{\bibinfo{volume}{A547}},
  \bibinfo{pages}{612} (\bibinfo{year}{1992}).

\bibitem[{\citenamefont{Zhitnitsky}(1994)}]{Zh94}
\bibinfo{author}{\bibfnamefont{A.~R.} \bibnamefont{Zhitnitsky}},
  \bibinfo{journal}{Phys. Lett.} \textbf{\bibinfo{volume}{B329}},
  \bibinfo{pages}{493} (\bibinfo{year}{1994}).

\bibitem[{\citenamefont{Gel'fand and Shilov}(1964)}]{GS64}
\bibinfo{author}{\bibfnamefont{I.~M.} \bibnamefont{Gel'fand}} \bibnamefont{and}
  \bibinfo{author}{\bibfnamefont{G.~E.} \bibnamefont{Shilov}},
  \emph{\bibinfo{title}{Generalized Functions}}, vol.~\bibinfo{volume}{1}
  (\bibinfo{publisher}{Academic}, \bibinfo{address}{New York},
  \bibinfo{year}{1964}).

\end{thebibliography}

\end{document}